\documentclass[11pt, letter]{article}

\usepackage{stolenstyle}

\usepackage{amsmath,epsf,amssymb,latexsym,amsthm,setspace,bbm,array,pifont}

\DeclareFontFamily{U}{rsf}{}
\DeclareFontShape{U}{rsf}{m}{n}{
  <5> <6> rsfs5 <7> <8> <9> rsfs7 <10-> rsfs10}{}
\DeclareMathAlphabet\Scr{U}{rsf}{m}{n}

\def\CO#1#2{{[#1,#2]}}
\def\AC#1#2{{\{#1,#2\}}}

\def\rep#1{{{\boldsymbol{#1}}}}

\def\C{{\mathbb C}}

\def\P{{\mathbb P}}
\def\R{{\mathbb R}}
\def\Z{{\mathbb Z}}
\def\N{{\mathbb N}}

\def\SO{\operatorname{SO}}

\def\SU{\operatorname{SU}}
\def\GU{\operatorname{U{}}}

\def\p{\partial}
\def\pb{\bar{\partial}}

\def\ff#1#2{{\textstyle\frac{#1}{#2}}}

\def\cF{{\cal F}}

\def\cK{{\cal K}}
\def\cL{{\cal L}}

\def\cO{{\cal O}}

\def\cT{{\cal T}}
\def\cU{{\cal U}}

\def\ep{{\epsilon}}

\newcommand\phib{\overline{\phi}}

\newcommand\ept{\widetilde{\ep}}

\newcommand\vphi{\varphi}

\newcommand\Thetab{\overline{\Theta}}

\newcommand\ub{\overline{u}}

\newcommand\At{\widetilde{A}}

\usepackage{graphicx}

\usepackage{tikz}

\usetikzlibrary[shapes.geometric,decorations.pathreplacing,automata,positioning]

\tikzset{surcirc/.style={shape=circle,inner sep=3pt,draw}}
\tikzset{surcirc2/.style={shape=circle,inner sep=1.5pt,draw}}

\def\TT{\mathcal{T}}

\def\bP{{\boldsymbol{P}}}

\def\bX{{\boldsymbol{X}}}

\def\W{{\mathcal{W}}}
\def\ds{\displaystyle}

\def\preprint{~~}

\title{Dynamics on asymptotically conical geometries}
\author[a] {Rebecca Field,}
\author[b] {Ilarion V.~Melnikov,}
\author[c] {and Bryce Weaver}
\affiliation[a]{Department of Mathematics and Statistics~}
\affiliation[b]{Department of Physics and Astronomy \\
James Madison University, Harrisonburg, VA 22807, USA}
\affiliation[c]{Department of Mathematics,\\
Xavier University, Cincinnati, OH 45207, USA}
\emailAdd{fieldre@jmu.edu}
\emailAdd{melnikix@jmu.edu}
\emailAdd{weaverb3@xavier.edu}

\abstract{We obtain general results on the dynamics of exactly conical geometries, where we use the notion of boundaries at infinity to characterize asymptotic behavior.  As we demonstrate in examples, these notions also  apply to smooth geometries that are merely asymptotically conical, such as the Eguchi-Hanson or resolved conifold geometries.  In these cases we obtain a rather complete qualitative understanding of the varieties of asymptotic behavior, and we probe the connectivity of the phase space by finding infinitely large families of multiple geodesics connecting a point on the infinite past boundary with a point in the infinite future boundary.}

\begin{document}

\maketitle

\section{Introduction} \label{s:intro}
The conifold~\cite{Candelas:1989jb,Candelas:1989js} has played a key role in string theory, in particular in stringy geometry and the AdS/CFT correspondence.  The ubiquity of this geometry is explained by its appearance as a local model of degeneration in many more complicated compact geometries.  Its utility is largely due to two features:  first, in the degeneration limit certain observables become largely insensitive to global features of the complicated compact geometry; second, the conifold geometry itself is remarkably simple and highly symmetric.  In short, one is able to obtain both simple and universal results.  Naturally, there are also many refinements, where one breaks the underlying symmetries, turns on backgrounds for fields other than the metric, or changes the topology altogether.  As we show below, it also gives a beautifully simple yet rich example of a classical dynamical system.  

Consider the resolved conifold as a solution to the string equations of motion that preserves $16$ (in the case of type II string) or $8$ (in the case of heterotic string) supercharges.  The conifold is an example of an asymptotically conical (AC) space, with a metric that asymptotes to
\begin{align}\label{e:alemetric}
\lim_{r/a \to \infty} g = dr^2 + r^2 g_{\text{link}}~.
\end{align}
Here $a$ is a parameter with dimensions of length---in our case the radius of the exceptional curve at $r=0$, and $g_{\text{link}}$ is a smooth metric on a $5$-dimensional compact ``link''manifold, in our case, the famous $T^{1,1}$ space.   Since this geometry has a well-understood asymptotic region, it is natural to consider string scattering on this non-trivially curved background.  In string perturbation theory this is obtained by computing correlation functions in the conformal field theory associated to the conifold geometry.  In this work, we examine the simplest aspect of this problem---the classical geodesics.  

We consider a localized excitation created at some large radial distance $r_0 \gg a$ with an initial radial motion towards the $r \approx 0$ region and some choice of initial momenta in the link (we will often say ``angular'') directions.  Most such initial data leads to geodesics that return again to $r\approx r_0$ in some time $2 T$.  To solve the scattering problem we must find $T$ and describe the motion on the link manifold.  

The large symmetry reduces the dynamics to a Hamiltonian for one degree of freedom---the radial coordinate---with an effective potential that depends on initial link data.  We use this description to obtain a complete solution to the scattering problem.  We find explicit formulas for the motion on the link manifold in terms of integrals for angular shifts, and we characterize the initial data corresponding to ``lost geodesics,''  which are solutions that asymptotically approach the exceptional curve and never escape to large $r$.  The existence of the latter solutions also suggests that we should have multiple geodesics between the same initial and final data, and we find infinite families of such multiple geodesics.

This rather complete understanding of the dynamics is a useful springboard for further exploration:  in future work we will obtain semi-classical results and compare them to exact SCFT descriptions of the sort originally proposed in~\cite{Ghoshal:1995wm} and explored further in~\cite{Giveon:1999zm,Eguchi:2004yi,Eguchi:2004ik,Ashok:2007ui,Mizoguchi:2008mk}.\footnote{ Another way to generalize our results is to consider more general classical geometries.  It has recently been shown that the dynamics on the $Y^{p,q}$ spaces is completely integrable~\cite{Babalic:2015fia,Visinescu:2016xna}, and it would be very interesting to extend our results to more general spaces that are resolutions of cones over $Y^{p,q}$; however, this may not be easy due to the reduced amount of symmetry and more complicated singularity structure.} 
For instance, in the semi-classical approximation the lost and the multiple geodesics are related to $\alpha'$--non-perturbative corrections to the geometry, and it will be interesting to  make the correspondence more precise.

It is well-known that the Ricci-flat metric on a Calabi-Yau space does not yield an exact solution to the string equations of motion, so our results only describe the classical $\alpha' \to 0$ limit of the solution.  Nevertheless, the results we obtain are still quite interesting, since they explicitly incorporate the classical dependence of the scattering on the resolution parameter, something that is not easy to achieve in the exact SCFT description.  We also work out a very similar and in many ways simpler problem of geodesics on Eguchi-Hanson space:  in this case we do not expect string corrections to the background geometry, so we can hope to match our results more quantitatively to the corresponding SCFT.

These results are also relevant to the general theory of dynamical systems.  One particular feature that we  explore is the existence of ``boundaries at infinity.''  These boundaries are known to exist for non-positive sectional curvature, see for example ~\cite{10.2307/1971373, 10.2307/1971303, Eberlein:1996de,KNIEPER2002453}.  Similar boundaries at infinity have been used in understanding the limiting structure of groups, Dirichlet problems, and in understanding measures, and counting periodic orbits, see, e.g.~\cite{sullivan1979density,patterson1976limit,zbMATH04080312,10.2307/120995}. How prevalent such structures are for more general dynamical systems is not well understood.  We demonstrate the existence of boundaries at infinity for manifolds whose metric is given as a cone.  The boundaries at infinity exist for both Eguchi-Hanson and conifold geometries and are similar in structure to the conic setting, but with several questions yet to be answered in the connectivity of infinite pasts and infinite futures. Our main interest in these boundaries at infinity is to help characterize asymptotic behavior and to understand these behaviors qualitatively.


The rest of the paper is organized as follows.  We begin with a brief review of geodesic flows and our conventions in section~\ref{s:review}.  We follow this up with a general solution of geodesic flows and a study of boundaries at infinity for conical (in general, singular) geometries in section~\ref{s:cones}. Next, in section~\ref{s:EH}, we study the dynamics on the Eguchi-Hanson space.    Finally, we tackle the conifold in section~\ref{s:conifold}.

\section{Geodesic and geometric review} \label{s:review}

If $M$ is a Riemannian manifold with metric $g$, then in local coordinates we denote the metric by $g=g_{ij}\;dx^i\;dx^j$.  The Lagrangian for geodesic flow is $L=\frac12g_{ij}\dot x^i\dot x^j$~, and the Hamiltonian is $H=\frac12g^{ij}p_ip_j$~, where $g^{ij}$ is the inverse metric and $p_i=g_{ij}\dot{x}^j$ are the conjugate momenta.  
A curve $\gamma:I
\to M$ is a geodesic if it minimizes
\begin{align}
E(\gamma) &= \int_I L(\gamma) dt~.
\end{align}  
In terms of the Hamiltonian, 
at a point $x\in M$ with conjugate momenta $p$, the equations for geodesic flow are  
\begin{align}
\label{eq:Heom}
\ds\dot{x}^k&=\frac{\p H}{\p p_k}=g^{kj}p_j~,&
\ds\dot{p}_k&=-\frac{\p H}{\p x^k}=-\frac12g^{ij}_{~,k}p_ip_j~.
\end{align}
We use the summation convention (repeated indices are summed) and denote differentiation with respect to coordinates by $f_{,i} = \frac{\p f}{\p x^i}$.

Killing vectors correspond to continuous isometries of the manifold and constrain geodesic flow.
Given a vector field $V = V^k \frac{\p}{\p x^k}$,
the Lie derivative of $g$ with respect to $V$ is
\begin{align}
\label{eq:Lieg}
(\cL_V g)_{ij} & = V^k g_{ij,k} + g_{ik} V^k_{~,j} + g_{kj} V^k_{~,i}~,
\end{align}
and $V$ is Killing with respect to $g$ iff $\cL_V g = 0$.
Killing vectors lead to conserved quantities:  if  $V=V^k(x)\frac{\p}{\p x^k}$ is Killing,
then $P=p_iV^i$ is conserved along the geodesic.  In local coordinates
\begin{align*}
\frac{d}{dt} \left( p_i V^i \right) = V^i \dot p_i + p_i V^i_{,k} \dot x^k = -\ff{1}{2} (\cL_V g)^{ij} p_i p_j = 0~,
\end{align*}
where the second equality follows from~(\ref{eq:Heom}) and~(\ref{eq:Lieg})~.
%

\subsection*{Dynamical notions}

Now let us recall some basic definitions and topics from the theory of smooth dynamical systems and Riemannian geometry.  We restrict our dynamical systems to actions of the additive groups and semi-groups $\N$, $\Z$, $\R_{\ge 0}$, or $\R$. If the action is of the additive group $\Z$ or $\R$, then we say that the dynamical system is {\em invertible}.  If we have an action $h : X \times \TT \rightarrow X$, where $X$ is the phase space and $\TT$ is group or semi-group, we use the notation of $h^t(x)$ to denote the action of $t\in\TT$ on $x \in X$.  If $X$ is a metric space with metric $d$, the {\em stable set} of $x \in X$ is 
\begin{align} 
		\W^s(x) &= \left\{y \in X : \lim_{t\rightarrow \infty } d\left(h^t(x),h^t(y)  \right) = 0 \right\}. 
\end{align} 
Furthermore, if the dynamical system is  invertible, the {\em unstable set} of $x$ is 
\begin{align} 
\W^u(x) = \left\{y \in X : \lim_{t\rightarrow -\infty } d\left(h^t(x),h^t(y)  \right)  = 0 \right\}. 
\end{align}

We now recall a few characteristics of geometry and dynamics on manifolds with non-positive sectional curvature.  As we will see in section~\ref{s:cones}, there are close parallels between geodesic flow on such manifolds with the flows on manifolds with conic metric.   

A manifold $M$ has non-positive sectional curvature if the sectional curvature satisfies $K(v_x,w_x) \le 0$ for any $v_x,w_x \in TM$.\footnote{For a thorough discussion on non-positive sectional curvature, see~\cite{Eberlein:1996de}.}  Here and in what follows we use a short-hand notation for vectors in the tangent bundle of a Riemannian manifold $M$:  the subscript denotes base point, e.g. $v_x \in T_xM$, where $x\in M$ is the base point.

We now review some aspects of the geodesic flow $g^t : TM \rightarrow TM$ for $M$ with  non-positive sectional curvature.   
A complete simply connected Riemannian manifold with non-positive sectional curvature is diffeomorphic to $\R^n$, where $n$ is the dimension.  This is equivalent to saying that the universal cover of any manifold with non-positive sectional curvature is $\R^n$. For the rest of this section, we consider $M$ to be the universal cover for a Riemannian manifold with non-positive sectional curvature. The metric $d$ that we use on the tangent bundle is the Sasaki metric~\cite{sasaki1958}, given in local  
coordinates by 
\begin{align} \label{e:sasaki} d\sigma^2 &= g_{ij}\;dx^i\;dx^j + g_{ij}\;Dv^i\;Dv^j~,
\end{align} 
where $Dv^i$ is the covariant differential
$$Dv^i = dv^i + \Gamma^i_{kl}v^kdx^l~,$$ 
and $\Gamma^i_{kl}$ are the Christoffel symbols.

Second, vectors $v_x, v_y \in T^1M$ in the unit tangent bundle can be given an equivalence relation where $v_x \sim v_y$ if $\lim_{t\rightarrow \infty} d(g^t(v_x), g^{t}(v_y)) \le C$ for some constant $C>0$ depending on $v_x, v_y$ (similar definition for $t\rightarrow -\infty$).   Let $\partial_{v_x}^{+\infty}$ be the equivalence class for $v_x$ (similar for $\partial_{v_x}^{-\infty}$).  These equivalence classes form the ``forward boundary at infinity'', $\partial^{+\infty}M$ (similar for backward boundary at infinity, $\partial^{-\infty}M$).  It is non-trivial, but for any $x\in M$, the map $T^1_xM \rightarrow \partial^{+\infty}M$ defined by $v_x \rightarrow \partial_{v_x}^{+\infty}$ is one-to-one (similar for  $\partial^{-\infty}M$).  Via such an identification, both boundaries at infinity can be given the topology of $S^{n-1}$.  By fixing a particular $x\in M$,  we have a one-to-one relationship between $S^{n-1} \times S^{n-1}\sim\partial^{-\infty}M \times \partial^{+\infty}M$ via the natural identification of both the former and the latter to $T^1_xM \times T^1_xM$.

Note that for any $v_y \in T^1M$, $g^t(v_y) \sim v_y$. To distinguish these time shifts of vectors, for $v_x \in T^1_xM$ where $x$ arbitrary but fixed, the {\em Busemann function} $f_{v_x} : M \rightarrow \R$ is defined by 
\begin{align} 
f_{v_x}(y) &= \lim_{t\rightarrow +\infty} d(y,\gamma_{v_x}(t)) - t~,
\end{align}
where $\gamma_{v_x}$ is the geodesic with initial condition $v_x$. The introduction of the Busemann function allows for a more complete understanding of the full the dynamical system as it distinguishes points that have the same asymptotic behavior via a standardization of time.    For existence\footnote{Its existence relies on $M$ being simply connected and having non-positive sectional curvature.} and  properties of the Busemann functions see~\cite{Eberlein:1996de}, page 45.  We use it to standardize time in the following way.  For any $v_y \in T^1M$, there is $v_x \in T^1_xM$ such that $\partial_{v_x}^{+\infty}= \partial_{v_y}^{+\infty}$.  We define the time for $v_y$ by $t_{v_y} = f_{v_x}(y)$.  This has the nice property that $t_{g^t(v_y)} = t + t_{v_y}$.  

For the identification of the geodesic flow, let 
\begin{align*}\left(S^{n-1} \times S^{n-1}\right)_{\mbox{ \scriptsize C}} &= \left\{ \left(w_x,v_x\right) \in S^{n-1} \times S^{n-1} \: \left| \:  w_x\sim_{\mbox{\scriptsize geo}} v_x  \right. \right\}~,\end{align*}  where $w_x\sim_{\mbox{\scriptsize geo}} v_x$
if there exists $v_z \in T^1M$ such that both $\partial_{v_z}^{-\infty} = \partial_{w_x}^{-\infty}$ and $\partial_{v_z}^{+\infty} = \partial_{v_x}^{+\infty}$.\footnote{This is sometimes referred as {\em connecting} the points in the boundaries at infinity.} All together, we have the map from the unit tangent bundle
\begin{align} 
P &: T^1M \rightarrow \left(S^{n-1} \times S^{n-1}\right)_{\mbox{ \scriptsize C}} \times \R
\end{align}
where  and $P(v_y) = (\partial_{v_y}^{-\infty},\partial_{v_y}^{+\infty}, f_{v_x}(y))$ with $\partial_{v_y}^{+\infty} = \partial_{v_x}^{+\infty}$.
Via this identification, the geodesic flow, $g^t$, satisfies the commutative diagram in figure~\ref{f:nonnegdiag}, where
\begin{align} 
\boxplus^t\left(\partial_{v_y}^{-\infty},\partial_{v_y}^{+\infty},f_{v_x}(y)\right) &= \left(\partial_{v_y}^{-\infty},\partial_{v_y}^{+\infty},f_{v_x}(y)+t\right)~. \label{e:boxy1}\end{align}

\begin{figure} \begin{center}
\begin{tikzpicture}[node distance=4.7cm, auto] 
  \node (A) {$T^1M$};
  \node (B) [right of=A] {$T^1M$};
  \node (Ab) [below of=A] {$\left(S^{n-1} \times S^{n-1}\right)_{\mbox{ \scriptsize C}} \times \R$};
  \node (Bb) [below of=B] {$\left(S^{n-1} \times S^{n-1}\right)_{\mbox{ \scriptsize C}} \times \R$};
  \draw[->] (A) to node {$g^t$} (B);
  \draw[->] (Ab) to node [swap] {$\boxplus^t$} (Bb);
  \draw[->] (A) to node   [swap] {$ P $} (Ab);
  \draw[->] (B) to node  {$ P $} (Bb);
\end{tikzpicture}
\caption{Commutative diagram for geodesics on manifolds with non-positive sectional curvature} \label{f:nonnegdiag}
\end{center} \end{figure}

The mapping $P$ is onto but is not, in general, one-to-one.  From an asymptotic dynamical systems perspective, if the mapping is not one-to-one the combined orbits have the same past and future and therefore do not represent distinct behavior.  This changes in the setting of bounded negative sectional curvature (defined by the existence of a $c>0$ such that $-c \le K(v_x,w_x) \le -1/c$ for any $v_x,w_x \in T_xM$ -- e.g. universal cover of a compact manifold with negative sectional curvature).    We can then, uniquely, characterize vectors $v_y$ in our geodesic flow via $P$. Furthermore, in this setting, $\left(S^{n-1} \times S^{n-1}\right)_{\mbox{ \scriptsize C}} = \left(S^{n-1} \times S^{n-1}\right) \setminus \Delta$ where $\Delta = \left\{(-v_x,v_x) \:|\: v_x \in S^{n-1} \right\}$.
\footnote{This is like the Poincare disc model of hyperbolic space.}

\section{Flows on the cone} \label{s:cones}
\subsection{Conic geometry}
We consider geodesics on a warped product of the form $\R_{> 0} \times N$ with metric
\begin{align}
ds^2 = dr^2 + f(r) h_{ij}(x) dx^i dx^j~.
\end{align}
Here, the $x^i$ are local coordinates on the link manifold $N$, and $h$ is the Riemannian metric on $N$.
We show that geodesics on the warped product project to geodesics on the link manifold $N$, and in the particular case of a cone, $f(r)=r^2$, the length of a projected complete geodesic is precisely $\pi$. 
The Lagrangian is 
\begin{align}
L = \frac{1}{2} \dot r^2 + \frac{1}{2} f(r) h_{ij} \dot x^i \dot x^j~,
\end{align}
and the momenta are therefore
\begin{align}
p_r & = \dot r~, &
p_i &= f(r) h_{ij} \dot x^j~. \label{e:momentums}
\end{align}
The corresponding Hamiltonian is 
\begin{align}
H & = \frac{1}{2} p_r^2 + \frac{1}{f(r)} H_N~, &  H_N = \frac{1}{2} p_i p_j h^{ij}~.
\end{align}

\subsection{Radial dynamics}
The radial dynamics are independent of the dynamics on the link manifold. To see this,
consider a particular level set $E$, so that
\begin{align}
E & = \frac{1}{2}p_r^2 + \frac{1}{f(r)} H_N = \text{constant}~, &\implies
H_N & = f(r) \left( E - \frac{1}{2} p_r^2 \right)~.
\end{align}
The $r$ equations of motion are then
\begin{align}
\dot r & = p_r ~, &
\dot p_r & = \frac{f'}{f^2}  H_N~.
\end{align}

This leads to two conclusions.  First, that $H_N$ is a constant of the motion:
\begin{align}
\dot H_N & = \frac{f'}{f}  \dot r H_N-f p_r \dot p_r =  \frac{f'}{f} p_r H_N - f p_r \left(\frac{f'}{f^2}  H_N\right) = 0~.
\end{align}
A more formal way to realize this is to compute the Poisson bracket:
\begin{align}
\dot H_N = \AC{H}{H_N} = 0~.
\end{align}
Note that this does not rely on any special properties of $H_N$ or $f$; \textit{any} $H_N$ and \textit{any} $f$ work.

Second, we can integrate the $r$:
\begin{align}
\dot r = p_r = \pm \sqrt{2E - \frac{ 2H_N}{f}}~.\label{eq:integratingr}
\end{align}
Indeed, the dynamics of $r$ is described by an effective Hamiltonian
\begin{align}
H_{\text{eff}} = \frac{1}{2} p_r^2 + \frac{C}{f(r)}~,
\end{align}
where $C = H_N \ge 0$ is a constant.  Therefore, all of the dynamics is determined by properties of $f(r)$.

\subsection{The projection to the link manifold}\label{ss:ptlm}
Consider the projection of a solution to the Hamiltonian equations to the link manifold. Knowing that $H_N$ is conserved, consider the change of variable $\tilde{t}$, where 
\begin{align}
\frac{dt}{d\tilde{t}} = f(r)~.
\end{align}
Using the same formulation for $H_N$, consider the equations of motion dictating the projection of an an integral curve  under this change of time (a priori, these equation contain $r$, hence are time dependent).
The equations for the motion become
\begin{align}
\frac{dx^i}{d\tilde{t}} &= \frac{dt}{d\tilde{t}} \dot{x}^i = f(r) \frac{\p { H}}{\p p_i} = f(r) \frac{1}{f(r)}\frac{\p { H_N}}{\p p_i} = \frac{\p { H_N}}{\p p_i}~, \\
\frac{dp_i}{d\tilde{t}} &= \frac{dt}{d\tilde{t}} \dot{p}_i = -f(r) \frac{\p { H}}{\p x^i} = -f(r) \frac{1}{f(r)}\frac{\p { H_N}}{\p x^i} =- \frac{\p { H_N}}{\p x^i}~,
\end{align}
from the equations above and standard formalisms.  
These, one readily notes, are the equations of motion for the the Hamiltonian $H_N$ on $N$.
  
\subsection{Turning points and properties for a cone, $f = r^2$} \label{ss:tpandprop}
An important special case, applicable to both Eguchi-Hanson and the conifold, is when $f = r^2$.  In this case, we can integrate for $r$ explicitly (\ref{eq:integratingr}).  The turning point $r=r_\ast$ is determined by
\begin{align}
\dot r^2 = 2E -\frac{2 H_N}{r^2} =0 \implies r_\ast = \sqrt{\frac{H_N}{E}}~.
\end{align}
Hence, we see that $r_* = 0$ is only possible if $H_N=0$.  We can manipulate this to obtain
\begin{align}
\frac{r \dot r}{\sqrt{r^2-r_\ast^2}} = \pm \sqrt{2E}~,
\end{align}
which integrates to
\begin{align}
r^2 = r_\ast^2 + 2E(t-t_\ast)^2~. \label{e:roft}
\end{align}
Without loss of generality, we can take the turning point time to be $t_\ast = 0$.\footnote{The use of the turning point $r_\ast$ to standardize time $0$ replaces the use of the Busemann function for non-positive sectional curvature geometries, see section~\ref{s:review}.}

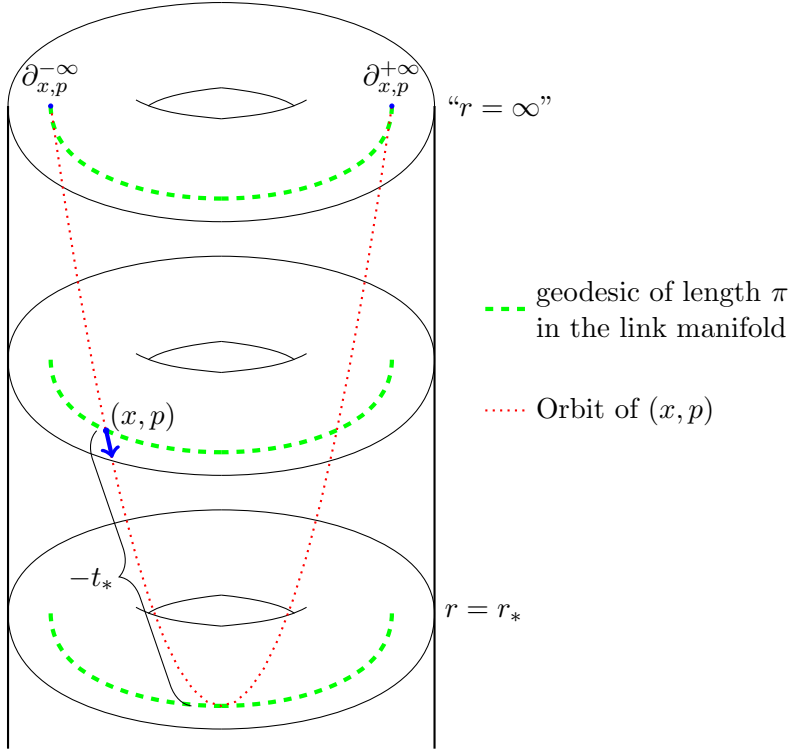
\begin{figure}
\begin{center}
\begin{tikzpicture}[scale=0.9]

\foreach \x in {0.9} \foreach \y in {0, 3.75, 7.5} \foreach \z in {0.8}
		{
		\draw (-3.5*\x,{(0*\x)+\y}) .. controls (-3.5*\x,{(1.2*\x)+\y}) and (-1.5*\x,{(1.7*\x)+\y}) .. (0*\x,{(1.7*\x)+\y});
		\draw (3.5*\x,{(0*\x)+\y}) .. controls (3.5*\x,{(1.2*\x)+\y}) and (1.5*\x,{(1.7*\x)+\y}) .. (0*\x,{(1.7*\x)+\y});
		\draw (-3.5*\x,{(0*\x)+\y}) .. controls (-3.5*\x,{(-1.4*\x)+\y}) and (-1.5*\x,{(-1.9*\x)+\y}) .. (0*\x,{(-1.9*\x)+\y});
		\draw (3.5*\x,{(0*\x)+\y}) .. controls (3.5*\x,{(-1.4*\x)+\y}) and (1.5*\x,{(-1.9*\x)+\y}) .. (0*\x,{(-1.9*\x)+\y});
		\draw (-1.4*\x,{(.1*\x)+\y}) .. controls (-1*\x,{(-0.12*\x)+\y}) and (-0.5*\x,{(-0.16*\x)+\y})  .. (0*\x,{(-.21*\x)+\y}) ;
		\draw (1.4*\x,{(.1*\x)+\y}) .. controls (1*\x,{(-0.12*\x)+\y}) and (0.5*\x,{(-0.16*\x)+\y})  .. (0*\x,{(-.21*\x)+\y}) ;
		\draw (-1.2*\x,{(0*\x)+\y}) .. controls (-1*\x,{(0.15*\x)+\y}) and (-0.5*\x,{(0.25*\x)+\y})  .. (0*\x,{(.3*\x)+\y}) ; 
		\draw (1.2*\x,{(0*\x)+\y}) .. controls (1*\x,{(0.15*\x)+\y}) and (0.5*\x,{(0.25*\x)+\y})  .. (0*\x,{(.3*\x)+\y}) ;
		\draw[dashed,ultra thick,green] (3.5*\x*\z,{(0*\x*\z)+\y}) .. controls (3.5*\x*\z,{(-1.4*\x*\z)+\y}) and (1.5*\x*\z,{(-1.9*\x*\z)+\y}) .. (0*\x*\z,{(-1.9*\x*\z)+\y}); 
		\draw[dashed,ultra thick,green] (-3.5*\x*\z,{(0*\x*\z)+\y}) .. controls (-3.5*\x*\z,{(-1.4*\x*\z)+\y}) and (-1.5*\x*\z,{(-1.9*\x*\z)+\y}) .. (0*\x*\z,{(-1.9*\x*\z)+\y});
		}
		
	\draw[thick,-] (3.5*0.9,-2) -- (3.5*0.9,7.5) ;
	\draw[thick,-] (-3.5*0.9,-2) -- (-3.5*0.9,7.5) ;
	
	\draw[red, dotted, thick, domain=-2.52:2.52] plot (\x, {((7.5 + (1.9*.8*0.9))/(3.5*.8*0.9)^2)*\x*\x - 1.9*.8*0.9 });
	\node[above] at (-3.5*.8*0.9, 7.5) {$\partial_{x,p}^{-\infty}$} ;
	\draw[blue,fill] (-3.5*.8*0.9, 7.5) circle [radius=.03];
	\node[above] at (3.5*.8*0.9, 7.5) {$\partial_{x,p}^{+\infty}$} ;
	\draw[blue,fill] (3.5*.8*0.9, 7.5) circle [radius=.03];
	
	\draw[blue,fill] (-1.7, 2.7) circle [radius=.04];
	\node[right] at (-1.8, 2.9) {$(x,p)$} ;
	\draw[blue,ultra thick,->] (-1.7, 2.7) -- (-1.61, 2.3) ;
	
	\draw [decorate,decoration={brace,amplitude=10pt},xshift=-4pt,yshift=0pt]
(-.3,- 1.9*.8*0.9) -- (-1.7, 2.7) node [black,midway,xshift=-0.7cm, yshift=-4pt] 
{ $-t_\ast$};
	
	\node[right] at (3.5*.9, 7.5) {``$r=\infty$''} ;
	\node[right] at (3.5*.9, 0) {$r=r_\ast$} ;
	
	\draw[dashed,ultra thick,green] (3.9, 4.5) -- (4.5, 4.5) ;
	\node[right, text width=3.5cm] at (4.5, 4.5) {geodesic of length $\pi$ in the link manifold} ;
	\draw[red, dotted, thick] (3.9, 3) -- (4.5, 3) ;
	\node[right, text width=3cm] at (4.5, 3) {Orbit of $(x,p)$} ;
\end{tikzpicture} \caption{Conic geodesics} \label{f:congeo}
\end{center}

\end{figure}

From this, we can compute the total distance traversed in the link manifold as follows.  First, for any $f(r)$, we have
\begin{align}
	S = \int ds = \int^{\infty}_{-\infty} dt \sqrt{h_{ij} \dot x^i \dot x^j} = \sqrt{2H_N} \int^{\infty}_{-\infty} \frac{dt}{f(r)}~.
\end{align}
In the case of $f =r ^2$, the integral is
\begin{align}
	S = \sqrt{2 H_N} \int^{\infty}_{-\infty} \frac{ dt}{r_\ast^2 + 2 E t^2} = \int^{\infty}_{-\infty} \frac{d\tau}{1+\tau^2} = \pi~.
\end{align}
 Combining these results with those from Section~\ref{ss:ptlm} (as well as the form of our Hamiltonians), we see that the projection to the link manifold results in a geodesic curve of total length $\pi$.  These geodesic curves have two limiting endpoints: one as $t \rightarrow -\infty$, and the other as $t \rightarrow \infty$. 
For a given $(x,p)$ momentum in the cotangent bundle, we use the notation $\partial_{x,p}^{-\infty}$ for the {\em infinite past}, and $\partial_{x,p}^{+\infty}$ for the {\em infinite future} given by these limits respectively. For a vector $(x,\dot{x})$ in the tangent bundle, $\partial_{x,\dot{x}}^{-\infty}$ and $\partial_{x,\dot{x}}^{+\infty}$  denote the corresponding limits.  The results of this section are illustrated in figure~\ref{f:congeo}.

\subsection{Stable (unstable) manifolds} \label{ss:stabman}

 Due the the symmetry of the dynamical system coming from the Hamiltonian, we only need to consider the stable manifolds.  Take then the initial conditions  $(\tilde{x},\dot{\tilde{x}}) = (r_x,x,\dot{r}_x,\dot{x})$. For characterizing long-term behaviors, the distance we use on $\R_{> 0} \times N$ comes from the unwarped product metric
\begin{align}
	ds^2 = dr^2 + h_{ij}(x) dx^i dx^j~.
\end{align}  
Note that this is not the metric that determines the geodesic flow, but as we will see, it allows us to naturally group asymptotic behaviors.\footnote{This allows us to replace the identifications we made in section section~\ref{s:review} of staying bounded distance apart for non-positive sectional curvature geometries by distance going to zero for cones; see the definition of the boundary at infinity.}
For the tangent bundle, the distance is the one coming from the corresponding Sasaki metric (see section~\ref{s:review}, (\ref{e:sasaki})).  We use $d$ to denote the distances on both the manifold $\R_{> 0} \times N$ and the tangent bundle.    

Ignoring geodesics that are tangential to the radial direction, from section~\ref{ss:tpandprop} the remaining geodesics have a basic form: their radial coordinate, $r$, has a turning point $r_\ast$ at some $t_\ast$ and increases to $r=\infty$ as $t \to\pm \infty$.  Furthermore, 
\begin{align} 
	\lim_{t\rightarrow \infty} \dot{r_x} &=\sqrt{2E}, & \lim_{t\rightarrow \infty} \dot{x}^i &= 0
\end{align} 
for all $i$, and the projection of the geodesic to the link manifold traces out a geodesic arc of length $\pi$ in the link.  From these facts, two trajectories in the tangent bundle, $(r_x(t), x(t),\dot r_x(t), \dot{x}(t))$ and  $(r_y(t), y(t),\dot r_y(t), \dot{y}(t))$, satisfy 
\begin{align}
	\lim_{t\rightarrow\infty} d\left((r_x(t), x(t),\dot r_x(t), \dot{x}(t)),(r_y(t), y(t),\dot r_y(t), \dot{y}(t))\right) &= 0~,
\end{align}
iff  (i) $\lim_{t\rightarrow \infty} r_x(t) - r_y(t) = 0$ and (ii) $\lim_{t\rightarrow \infty} d_N(x(t),y(t)) =0$, where $d_N$ is the induced distance from the metric on the link manifold.  

From the end of section~\ref{ss:ptlm}, we have $\lim_{t\rightarrow \infty}  x(t) = \partial_{\tilde{x}_0,\dot{\tilde{x}}_0}^{+\infty}$, where $\tilde{x}_0 =(r_x(0),x(0))$ and  $\dot{\tilde{x}}_0 = (\dot{r}(0), \dot{x}(0))$.  Since a similar limit holds for $y(t)$,  (ii) occurs iff the trajectories have the same infinite future.  For (i), let $E_i$, $r_{i\ast}$, and  $t_{i\ast}$ represent the energy, turning point, and turning time respectively for  $i = x, y$.  From (\ref{e:roft}), with simplifications we have 
\begin{align}
	r_x(t) - r_y(t) &= \sqrt{r_{x\ast}^2 + 2E_x(t-t_{x\ast})^2} - \sqrt{r_{y\ast}^2 + 2E_y(t-t_{y\ast})^2} \nonumber \\
	&= \frac{2(E_x-E_y)t^2 + 4(E_xt_{x\ast}-E_yt_{y\ast})t + 2(E_xt^2_{x\ast}-E_yt^2_{y\ast})+ r_{x\ast}^2 - r_{y\ast}^2}{\sqrt{r_{x\ast}^2 + 2E_x(t-t_{x\ast})^2} + 					\sqrt{r_{y\ast}^2 + 2E_y(t-t_{y\ast})^2}}~.
\end{align}
This immediately yields that $\lim_{t\rightarrow \infty} r_x(t) - r_y(t) = 0$ if and only if the energies and turning times satisfy $E_x = E_y$ (obviously) and  $t_{x\ast}=t_{y\ast}$.

The stable manifolds in the tangent bundle can be characterized fairly easily for vectors at the turning point in terms of position and velocity.  This is given by 
\begin{align}
\W^s(r_\ast, x_0, 0, \dot{x}_0) &= \left\{(r, x, 0, \frac{\sqrt{2E}}{r}\dot{x}) \: \left| \: r \in \R_{>0}, x \in S_{\pi/2}\left(\partial^{+\infty}_{\left(r_\ast, x_0, 0, \dot{x}_0\right)}\right)  \right.\right\}~,
\end{align}
where $\partial^{+\infty}_{\left(r_\ast, x_0, 0, \dot{x}_0\right)}$ is the infinite future of $\left(r_\ast, x_0, 0, \dot{x}_0\right)$, $S_{\pi/2}$ is the sphere of radius $\pi/2$ about a base point in the link manifold, $\dot{x}$ is the inward pointing unit vector orthogonal to $S_{\pi/2}\left(\partial^{+\infty}_{\left(r_\ast, x_0, 0, \dot{x}_0\right)}\right)$ in $N$, and $E = \frac{r_{\ast}^2}{2} h_{ij} \dot{x}_0^i\dot{x}_0^j$.

\subsection{The boundaries at $\infty$ and relation to structure of the dynamical system} \label{ss:boundary}

From this development, we can also observe that the boundaries at infinity, $\partial^{-\infty}$ and $\partial^{+\infty}$, for the cone, can be identified with the link manifold. These boundaries at infinity capture  all of the interesting forward and backward asymptotic behaviors respectively.  To see how this helps understand asymptotic behaviors, consider the following construction.
The unit tangent bundle of the warped product can be mapped \begin{align} P: T^1\left(\R^+ \times N\right) \setminus T^1_{\mbox{\scriptsize rad}}\rightarrow  \left(N\times N\right)_\pi \times \R^+ \times \R~,\end{align} where $\left(N\times N\right)_\pi = \left\{ (x,y) \in N \times N \:|\: x\sim_\pi y \right\}$, $x\sim_\pi y$ if $x$ and $y$ can be connected via a geodesic of length $\pi$, $T^1_{\mbox{\scriptsize rad}}$ denotes vectors tangent to the radial direction, and $P\left(x,\dot{x}\right) = \left(\partial_{x,\dot{x}}^{-\infty},\partial_{x,\dot{x}}^{+\infty},r_\ast,-t_\ast\right)$.  Via this identification, the geodesic flow, $g^t$, satisfies the commutative diagram found in figure~\ref{f:conediag}, where
\begin{align} 
\boxplus^t\left(\partial_{x,\dot{x}}^{-\infty},\partial_{x,\dot{x}}^{+\infty},r_\ast,-t_\ast\right) &= \left(\partial_{x,\dot{x}}^{-\infty},\partial_{x,\dot{x}}^{+\infty},r_\ast,-t_\ast +t\right)~. \label{e:boxy}\end{align}
 
\begin{figure} \begin{center}
\begin{tikzpicture}[node distance=4.7cm, auto] 
  \node (A) {$T^1\left(\R^+ \times N\right) \setminus T^1_{\mbox{\scriptsize rad}}$};
  \node (B) [right of=A] {$T^1\left(\R^+ \times N\right) \setminus T^1_{\mbox{\scriptsize rad}}$};
  \node (Ab) [below of=A] {$\left(N\times N\right)_\pi \times \R^+ \times \R$};
  \node (Bb) [below of=B] {$\left(N\times N\right)_\pi \times \R^+ \times \R$};
  \draw[->] (A) to node {$g^t$} (B);
  \draw[->] (Ab) to node [swap] {$\boxplus^t$} (Bb);
  \draw[->] (A) to node   [swap] {$ P $} (Ab);
  \draw[->] (B) to node  {$ P $} (Bb);
\end{tikzpicture}
\caption{Commutative diagram for geodesics on cones} \label{f:conediag}
\end{center}
\end{figure}

This mapping is onto, but is not one-to-one if there are conjugate points on the link manifold that are distance of $\pi$ apart.   
If the injectivity radius for the link manifold is greater than $\pi$, this becomes one-to-one.  We can then characterize vectors $(x,\dot{x})$ in our geodesic flow uniquely via their infinite past, $\partial_{x,\dot{x}}^{-\infty}$, infinite future, $\partial_{x,\dot{x}}^{+\infty}$, turning point, $r_\ast$, and negative the turning time, $-t_\ast$.  

\subsection{Comparison to non-positive sectional curvature}

The commutative diagrams in figures~\ref{f:nonnegdiag} and ~\ref{f:conediag} make clear there are some relationships between the structure of the asymptotic behaviors for the geodesic flow in the non-positive sectional curvature and conic settings. To illustrate the strength of these parallels, consider a case where the conic setting overlaps with non-positive sectional curvature: Euclidean space. Here the metric singularity at $r=0$ is non-essential, the link manifold is just the unit $S^{n-1}$, and $(x,y) \in \left(N\times N\right)_\pi$ if and only if $x$ and $y$ are polar opposites on the sphere.  The identification with  non-positive curvature is immediate:  two unit vectors in Euclidean space stay bounded forward distance apart if and only if they are going in the same direction (at the same speed, in this case 1).  Such vectors also would stay a bounded distance apart in negative time.  Therefore, in both settings, an infinite past is identified with a unique infinite future. 

There are differences in the boundaries at infinity between the conic and non-positive sectional curvature settings that are (obviously) not illustrated by the Euclidean space case.  First, the topology of the boundaries is always the link manifold for cones and is always $S^{n-1}$ for non-positive sectional curvature.  These just happened to coincide in the Euclidean case.  Second, the set of infinite pasts that connect to an infinite future in the Euclidean setting (one polar opposite point)  is atypical behavior for both  cones and for non-positive sectional curvature. For the standard behavior in cones, this set of infinite pasts is a co-dimension one sphere in the link manifold.  This contrasts, in particular, with the negative sectional curvature setting: any point on the forward boundary can be reached from any point on the negative boundary, except the one identified with itself.  

\section{Geodesics on the Eguchi-Hanson space} \label{s:EH}
\subsection{Eguchi-Hanson geometry}
The Eguchi-Hanson ALE space is quite well-known~\cite{Eguchi:1978xp}; it is the simplest member in a large class of four-dimensional ALE geometries~\cite{Gibbons:1979zt} that may be obtained as hyper-K\"ahler quotients~\cite{MR992334}. A  lucid review of general ALE geometry is given in~\cite{Joyce:2000cm}.  The geometry is a hyper-K\"ahler metric
on the non-compact manifold $M = \cO(-2) \to \P^1$.  Using $u$ for an affine coordinate on the base and $\phi$ for the coordinate on the fiber, the metric is derived from the K\"ahler potential
\begin{align}
\cK & = \sqrt{16a^4+R} + \ff{1}{2} \log \frac{ \sqrt{16a^4+R}-1}{\sqrt{16a^4+R}+1}~, &
R & = 4 \phi\phib (1+ u\ub)^2~.
\end{align}
Here $a$ is the radius of the base $\P^1$, as may be seen by writing the Hermitian form derived from $\cK$:
\begin{align}
\omega &= i \p\pb \cK = \frac{iR}{4 \sqrt{16a^4+R}} \Theta\wedge\Thetab + \sqrt{16a^4+R} \frac{ i du d\ub}{(1+u\ub)^2}~, &
\Theta & = \frac{d\phi}{\phi}+\frac{2 \ub du}{1+u\ub}~.
\end{align}
When $\phi \to 0$ the second term in $\omega$ reduces to the standard Fubini-Study metric on $\P^1$ with radius $a$.  Taking the limit $a\to 0$, we 
find that the exceptional divisor shrinks, and the space reduces to the orbifold $\C^2/\Z_2$.

The metric has a $\GU(2)$ isometry group, and it is convenient to choose coordinates that make this isometry manifest and also clearly
describe the asymptotic region.  New real coordinates, $r, \theta, \vphi,\psi$, are obtained by setting
\begin{align}
u & = \frac{\cos\theta-1}{\sin\theta} e^{-i\vphi}~, &
\phi & = \frac{(1+\cos\theta)r^2}{4} e^{i(\psi+\vphi)}~.
\end{align}
Here $r$ is a radial coordinate; $\theta$ and $\vphi$ are standard angular coordinates on $S^2$ with $0\le \theta \le \pi$ and $0\le \vphi < 2\pi$, while $\psi$ has
the range $0 \le \psi < 2\pi$, with $\psi \sim \psi+2\pi$.  In these coordinates the metric takes the form
\begin{align}
\label{eq:EHgeneralmetric}
g & = \frac{dr^2}{F(r)} + \frac{r^2}{4F(r)} \Psi^2 + \frac{r^2 F(r)}{4} \left(d\theta^2+\sin^2\theta d\vphi^2\right)~, 
\end{align}
where
\begin{align}
\Psi & = d\psi + \cos\theta d\vphi~,&
F(r) & = \sqrt{1+ \frac{16a^4}{r^4}}~.
\end{align}
The metric has a coordinate singularity at $r= 0$; this can be put in a more familiar form by setting 
\begin{align}
\label{eq:EHorigincoordinates}
x+iy = \frac{r^2}{4a} e^{i\psi}~.
\end{align}
%
Near $x,y = 0$ the 
metric takes the form
\begin{align}
g =  \left( dx^2+dy^2\right) + 2\cos\theta d\vphi (xdy-ydx) + a^2 (d\theta^2+\sin^2\theta d\vphi^2) + O(x^2+y^2)~.
\end{align}
So, the metric is indeed smooth and reduces to the expected form at the exceptional divisor (i.e $x=y=0$ in these coordinates).
This also offers a check of the periodicity of the angular coordinate $\psi$.

On the other hand, for large $r$ the metric is asymptotic to the conical metric obtained by setting $a=0$:
\begin{align}
g_{\text{cone}} &  = dr^2 + \frac{r^2}{4} \Psi^2 + \frac{r^2}{4} (d\theta^2+\sin^2\theta d\vphi^2)~, \nonumber\\
 & = dr^2 + \frac{r^2}{4} g_{\text{link}}~,
\end{align}
where $g_{\text{link}}$ is the metric on the link manifold $N$:
\begin{align}
g_{\text{link}} = d\theta^2+d\vphi^2+d\psi^2+2\cos\theta d\vphi d\psi~.
\end{align}
If $\psi$ had periodicity $4\pi$, this would be the round metric on $S^3$. Since $\psi$ has periodicity $2\pi$, we
see that our link is the lens space $N = S^3/\Z_2$.  This is, of course, not surprising given that $M$ is an ALE resolution of $\C^2/\Z_2$.


We now discuss the Killing vector fields.   First, we observe that the metric is independent of the coordinates $\vphi$ and $\psi$;  therefore, $\frac{\p}{\p\psi}$ and $V_3 = \frac{\p}{\p\vphi}$ are Killing vector fields.  We also have
\begin{align}
V_1 & = -\sin\vphi \frac{\p}{\p\vphi} - \cot\theta \cos\vphi \frac{\p}{\p\vphi} + \frac{\cos\vphi}{\sin\theta}\frac{\p}{\p\psi}~,&
V_2 &= \cos\vphi \frac{\p}{\p\theta} -\cot\theta\sin\vphi \frac{\p}{\p\vphi} +\frac{\sin\vphi}{\sin\theta} \frac{\p}{\p\psi}~.
\end{align}
This is easily derived by observing that these vector fields annihilate the standard round metric on $S^2$, as well as the $1$-form $\Psi$.  The $V_1$, $V_2$, and $V_3$ satisfy the $\SU(2)$ algebra:
\begin{align*}
\CO{V_A}{V_B} = -\ep_{ABC} V_C~.
\end{align*}
Finally, to construct the Hamiltonian, we need the inverse metric, and it is given by
\begin{align}
\label{eq:gEHinverse}
g^{-1} = F \left(\frac{\p}{\p r}\right)^2 + \frac{4}{r^2 F} \left(\frac{\p}{\p \theta}\right)^2
+ \frac{4}{r^2 F\sin^2\theta} \left(\frac{\p}{\p\vphi} - \cos\theta\frac{\p}{\p\psi}\right)^2
+ \frac{4F}{r^2} \left(\frac{\p}{\p\psi}\right)^2~.
\end{align}

\subsection{Hamiltonian and Conserved charges}
We now turn to the Hamiltonian for geodesic flow.  Applying the general formulas of section~\ref{s:review} to the Eguchi-Hanson metric,
we obtain
\begin{align}
H = \frac{1}{2} p_i p_j g^{ij}(x)
& = \frac{F}{2} p_r^2 +\frac{2 F}{r^2} p_{\psi}^2+ \frac{2}{r^2 F} \left(p_{\theta}^2 
+\frac{(p_{\vphi}-\cos\theta p_{\psi})^2}{ \sin^2\theta}\right)
~.
\end{align}

The total energy is, of course, preserved by geodesic flow; additional conserved quantities may be obtained from the Killing vectors (or more general Killing tensors) as reviewed in section~\ref{s:review}.  Using the $V_A$ and $\frac{\p}{\p\psi}$ we find the conserved quantities
\begin{align}
P_{\psi} &= p_{\psi}~,\qquad  P_{\vphi}  = p_{\vphi}~,  \nonumber\\
P_1 & = - p_{\theta} \sin\vphi +\frac{\cos\vphi}{\sin\theta} \left(p_{\psi} - \cos\theta p_{\vphi}\right)~, \nonumber\\
P_2 & = ~p_{\theta}\cos\vphi +\frac{\sin\vphi}{\sin\theta} \left(p_{\psi} - \cos\theta p_{\vphi}\right)~.
\end{align}
These expressions have  a simple geometric significance in terms of 
\begin{align}
\bP & = \begin{pmatrix} P_1 \\ P_2 \\ P_{\vphi} \end{pmatrix}~, &
\bX & = \begin{pmatrix} \cos\vphi \sin\theta \\ \sin\vphi \sin\theta \\  \cos\theta \end{pmatrix}~.
\end{align}
$\bX$ labels a point on the $S^2$ parametrized by $(\theta,\vphi)$, and $\bP$ satisfies
\begin{align}
\bP \cdot \bX & = P_{\psi}~.
\end{align}
This means the dynamics in $S^2$ is quite simple:  the unit vector $\bX$ precesses around the fixed vector $\bP$, as illustrated in the figure below.
\begin{align*}
\begin{tikzpicture}[>=stealth,scale=0.6]
\draw[thick] (0,0) circle [radius = 3];
\draw[thick,dashed] (0,0) ellipse [x radius = 3, y radius = 1];
\draw[thick] (-3,0) arc [start angle = 180, end angle = 360, x radius = 3, y radius = 1];
\draw[ultra thick, -> ] (0,0) -- (90:3.02) node [anchor=south] {$\bP$};
\draw[ultra thick, gray,->] (0,0) -- (50:2.1) node [anchor= north west] {$\bX$};
\draw[thin, dashed] (0,2) ellipse [x radius = 2.2, y radius = 0.5];
\draw (-2.2,2) arc [start angle = 180, end angle = 360, x radius = 2.2, y radius = 0.5];
\draw (0,0) node {$\bullet$};
\end{tikzpicture}
\end{align*}
It is not hard to check the Poisson brackets
\begin{align}
\AC{P_A}{P_B} & = \ep_{ABC} P_C~,&
\AC{P_A}{X_B} & = \ep_{ABC} X_C~.
\end{align}
Since the $P_A$ are generators of the $\SO(3)$ rotations, we see that, indeed, $\bP\cdot \bX$ is $\SO(3)$--invariant, and, thus, for any initial data
$\bX(0)$ and $\bP$, we can choose coordinates so that $P_1 = P_2 = 0$ and $P_\psi = P_{\vphi} \cos\theta$.  These relations imply, as the figure suggests, that $\theta$ is constant, and $p_{\theta} = 0$.\footnote{A short computation shows that with these constraints $\dot p_{\theta} = 0$, so
that the restriction to $p_{\theta} = 0$ is indeed consistent.}  Note that while $P_{\psi}$ is $\SO(3)$--invariant, the coordinate $\psi$ does
transform by a linear shift.

\subsection{Properties of geodesics}\label{ss:EHgeodesics}
To describe the properties of geodesics, we assume a coordinate choice as described in the previous section:  $\theta$ is constant; $p_{\theta} = 0$,
and $p_{\psi} = p_{\vphi} \cos\theta$.  Since $p_{\vphi} = P_{\vphi}$ is conserved, we find the following equations of motion for the cyclic coordinates:
\begin{align}
\label{eq:cycliceom}
\dot\vphi & = \frac{4 P_{\vphi}}{r^2 F}~, &
\dot\psi  & = \frac{ 4 P_{\vphi}\cos\theta}{r^2 F} (F^2 -1)~.
\end{align}
The dynamics of the radial coordinate $r$ is then captured by the effective Hamiltonian
\begin{align}
H_{\text{eff}} & = \frac{F}{2} p_r^2 + U_{\text{eff}}~, 
\end{align}
where the potential $U_{\text{eff}}$ is just a function of $r$ and the conserved quantities $P_{\vphi}$ and $\cos\theta$:
\begin{align}
U_{\text{eff}}(r) & = \frac{2 \sin^2\theta P_{\vphi}^2}{r^2 F} + \frac{2 F}{r^2} P_{\vphi}^2 \cos^2\theta = \frac{2P_{\vphi}^2}{r^2 F} \left(1  + \frac{16 a^4\cos^2\theta}{r^4}\right)~.
\end{align}
We then have
\begin{align}
\dot r & = F p_r~,
\end{align}
and conservation of energy determines
\begin{align}
p_r & = \pm \sqrt{\frac{2 (E-U_{\text{eff}})}{F}}~.
\end{align}

The qualitative dynamics are easily determined from $U_{\text{eff}}(r)$.  Suppose we consider a trajectory with $r(0) = r_0 >0$.
If we also take $p_r(0) >0$, then the flow takes us to large $r$, with $\dot r $ approaching a positive constant.  On the other
hand, if $p_r(0) <0$, then there are two basic scenarios.  There may be a turning point $r= r_\ast < r_0$ where $\dot r$ will vanish
and then change sign.  When $r$ is small, $U_{\text{eff}} \sim P_{\psi}^2 r^{-4}$ and, as long as $P_{\psi} \neq 0$, there will be a turning
point for any energy $E$.  If $P_{\psi} = 0$, then $U_{\text{eff}}(0) = P_{\vphi}^2/2a^2$, and there will be a turning point for any $E$
with $E < P_{\vphi}^2/2a^2$.  On the other hand, if $E\ge P_{\vphi}^2/2a^2$ we need to be more careful because of the coordinate
singularity at $r=0$.  As we show below, this does not pose any great difficulty.

\subsubsection*{The large $r$ behavior}
Consider first the Hamiltonian at large $r$.   Expanding for $r\gg 2a$, we obtain
\begin{align}
\dot \vphi &= \frac{4 P_{\vphi}}{r^2} + O\left(\left(\ff{2a}{r}\right)^6\right)~, &
\dot \psi & = O\left(\left(\ff{2a}{r}\right)^6\right)~,  \nonumber\\
\dot r & = \pm \sqrt{2E} \mp \frac{\sqrt{2}P_{\vphi}^2}{r^2\sqrt{E}} + O\left(\left(\ff{2a}{r}\right)^4\right)~.
\end{align}
For ``outward'' geodesics with $p_r(0)>0$  $\dot r$  approaches the constant value $\sqrt{2E}$, while for large $t$, both $\vphi$ and $\psi$ asymptote to constant values $\vphi_{\infty}$ and $\psi_{\infty}$: $\vphi = \vphi_{\infty} +O(t^{-1})$ and $\psi = \psi_{\infty} + O(t^{-5})$~.  Recall that in this choice of coordinate system, $\theta = \theta_{\infty}$ is fixed.

\subsubsection*{The small $r$ behavior}
When $r \ll 2a$, we must worry about the coordinate singularity at $r=0$.  Fortunately, this is easily handled by working in terms of the coordinates $x,y$ defined in~(\ref{eq:EHorigincoordinates}).   Since
\begin{align}
\label{eq:xymomenta}
p_r &= \frac{\p x}{\p r} p_x + \frac{\p y}{\p r} p_y = \frac{2}{r} \left(x p_x + y p_y\right)~, \nonumber\\
p_\psi &= \frac{\p x}{\p \psi} p_x + \frac{\p y}{\p \psi} p_y = x p_y - y p_x~,
\end{align}
it is easy to rewrite the Hamiltonian in these $x,y,\theta,\vphi$ coordinates.  We observe
\begin{align}
r^2 F = 4a^2\sqrt{1 + \frac{x^2+y^2}{a^2}}~,
\end{align}
so that
\begin{align}
\frac{F}{2} p_r^2 + \frac{2F}{r^2} p_\psi^2 = \frac{1}{2} (p_x^2+p_y^2)\sqrt{1 + \frac{x^2+y^2}{a^2}}~.
\end{align}
Combining that with the remaining terms in the Hamiltonian, we obtain
\begin{align}
H  =  \frac{1}{2} (p_x^2+p_y^2)\sqrt{1 + \frac{x^2+y^2}{a^2}} + \frac{1}{2a^2}
\left( p_\theta^2 + \frac{ (p_{\vphi}-\cos\theta (xp_y-yp_x))^2}{\sin^2\theta} \right)\frac{1}{\sqrt{1 + \frac{x^2+y^2}{a^2}}}~.
\end{align}
This makes two points quite clear:
\begin{enumerate}  
\item there are trapped geodesics on the exceptional divisor:  initial conditions with $x=y=0$ and $p_x = p_y=0$ are preserved by the flow and correspond precisely to geodesics on an $S^2$ of radius $a$;
\item the geodesics that pass through $r=0$ simply pass through the origin $x=y=0$, and emerge unscathed with $\psi \to \psi+\pi$.   
\end{enumerate}
We now make the second statement more precise.  Suppose we have a geodesic that passes through the origin.  Any such geodesic requires $p_{\psi} = xp_y-yp_x =0$ and energy $E > U_{\text{eff}}(0)$.  When $p_{\psi} = 0$, $U_{\text{eff}}(0) = P_{\vphi}^2/2a^2$, so that as $r\to 0$ we have
\begin{align}
p_r =-\frac{r}{2a} \sqrt{2E - \frac{P_{\vphi}^2}{a^2}} + O(r^5)~.
\end{align}
Since we also have $\dot \psi = 0$, $\psi$ is a constant, and we can solve for $p_x$ and $p_y$ by using~(\ref{eq:xymomenta}):
\begin{align}
p_x & = - \cos\psi \sqrt{2E - \frac{P_{\vphi}^2}{a^2}} + O(x^2+y^2)~,&
p_y & = - \sin\psi  \sqrt{2E - \frac{P_{\vphi}^2}{a^2}} + O(x^2+y^2)~.
\end{align}
So, the geodesic passes through the origin and is mapped, in terms of the $r,\psi$ coordinates, to an outward geodesic with
\begin{align}
p_r = +\frac{r}{2a} \sqrt{2E - \frac{P_{\vphi}^2}{a^2}} + O(r^5)~,
\end{align}
and the angle $\psi$ shifted as $\psi \to \psi +\pi$.

\subsubsection*{The lost geodesics:  $P_{\psi} = 0$ and $E = P_{\vphi}^2/2a^2$}
If a geodesic can approach $r=0$, and the energy is just tuned to the threshold value $E= P_{\vphi}^2/2a^2$, then the time to reach $r=0$ must diverge.  This is clear from basic properties of existence and uniqueness of geodesics and the existence of the trapped geodesics on the exceptional divisor; however, we can also see it explicitly.  We 
expand $Fp_r$ for $P_{\psi} = 0$ and $r \ll a$:
\begin{align}
F p_r = -\sqrt{2F\left(\frac{P_{\vphi}^2}{2a^2}-U_{\text{eff}}\right)} = -\frac{r}{2a} \sqrt{\frac{P_{\vphi}^2}{2a^2}} + O(r^5)~.
\end{align}
Thus,
\begin{align}
T_\ep = - \int^{r_0}_\ep \frac{dr}{F p_r}
\end{align}
has a logarithmic divergence as $\ep \to 0$.

In fact, it is easy to see that any lost geodesic is asymptotic to a trapped geodesic.  Given a point $x$ in the phase space satisfying our two lost geodesic conditions, the fact that $P_\psi=P_\vphi\cos\theta$ for our choice of coordinates implies that $\theta=\pi/2$.  Therefore, as $r\to 0$ and $T\to\infty$, the geodesic is asymptotic to the path around the equator in the exceptional curve.\footnote{In other words, there exists an axis with respect to which the geodesic is asymptotic to the equator.  We also see in the next section that just like time, the $\vphi$ coordinate exhibits logarithmic divergence, so the geodesic is asymptotic to the entire path along the equator.}

Although the existence of these ``lost geodesics'' is not surprising given the trapped geodesics on the exceptional divisor, they are certainly interesting from the point of view of an observer at large $r$.  Note that they occur in positive co-dimension in the phase space:  we need to set both $P_{\psi} = 0$ and $E = {P_{\vphi}^2}/{2a^2}$.

\subsubsection*{Radial geodesics: $P_{\vphi} = 0$}
When we set $P_{\vphi}= 0$, the angular coordinates all remain constant along the flow, and since
$U_{\text{eff}}$ also vanishes, the radial dynamics simply reduce to 
\begin{align}
\dot r & = F p_r~, & p_r = -\sqrt{2E/F}~.
\end{align}
We choose the minus sign in the expression for $p_r$ because we are only interested in geodesics that first approach $r=0$.  If we start with $r(0) = r_0$ and $p_r(0) = -\sqrt{2E/F(r_0)}$, then the trajectory  reaches $r =0$ at time $T$ given by
\begin{align}
\sqrt{2E} T & = \int_0^{r_0} \frac{dr} {\left(1 + \frac{16 a^4}{r^4} \right)^{1/4}}~.
\end{align}
Setting $x=(r/r_0)^{4}$, we can reduce the integral to a hypergeometric function:
\begin{align}
T & = \frac{r_0^2}{4a \sqrt{2E} } ~_2F_1\left((\ff{1}{4},\ff{1}{2};\ff{3}{2};-\left(\ff{r_0}{2a}\right)^4\right).
\end{align}
As discussed above, the geodesics passes  through the $r=0$ coordinate singularity with no drama:  $p_r$ switches sign and $\psi$  acquires
a shift of $\pi$.  It then takes the same amount of time $T$ to return to $r= r_0$.  So, we can summarize the
evolution as:
\begin{align}
\underbrace{(r_0,\theta_0,\phi_0,\psi_0)}_{t=0}  \to  \underbrace{(r_0,\theta_0,\phi_0,\psi_0+\pi)}_{t=2T}~.
\end{align}

\subsubsection*{Angular shifts}
For general momenta, the angular coordinates $\vphi$ and $\psi$ are no longer constant, but it is not too hard to write down
the integrals for the resulting shifts.  Suppose we have a trajectory starting with $r(0) = r_0$ and $p_r<0$, and let us assume that it either
reaches a turning point $r_\ast >0$ or reaches the origin $r_\ast = 0$ in finite time $T$.  From the discussion above the condition for this is 
simply $E \neq U_{\text{eff}}(0)$.  We then have the integral for $T$:
\begin{align}
T = \int_{r_\ast}^{r_0} \frac{ dr}{\sqrt{2F(E-U_{\text{eff}})}}~.
\end{align}
Using~(\ref{eq:cycliceom}) we also have
\begin{align}
\Delta \vphi = \vphi(T) - \vphi(0) = 4P_{\vphi}\int^T_0 \frac{dt}{r^2F} = 4 P_{\vphi} \int_{r_\ast}^{r_0} \frac{dr}{r^2 F\sqrt{2F(E-U_{\text{eff}})}}~,
\end{align}
and similarly
\begin{align}
\Delta \psi = \psi(T) -\psi(0) = 4 P_{\vphi}\cos\theta \int_{r_\ast}^{r_0} \frac{dr (F^2-1)}{r^2 F\sqrt{2F(E-U_{\text{eff}})}}~. 
\end{align}
A moment's thought shows that these integrals converge in all cases except those of the lost geodesics.  Away from the threshold at $E = U_{\text{eff}}(0)$,
the integral for $\Delta \vphi$ is obviously convergent.  The factor of $F^2-1$ in the integrand for $\Delta \psi$ scales as $1/r^4$; but this can only
lead to trouble when $r_\ast = 0$, and that requires $\cos\theta = 0$.  Moreover, although $T$ diverges when $r_0 \to \infty$, the angular shifts remain
finite.

\subsubsection*{Asymptotic angular shifts}
The integrals for the angular shifts simplify when $r_0 \to \infty$ and $r_\ast = 0$.    In fact, we can then ask for
the total angular shifts, i.e. the limit
\begin{align}
(\Delta\vphi)_{\infty} &= 2 \lim_{r_0 \to \infty} (\vphi(T) -\vphi(0))~.
\end{align}
This is the total change in the angle for a trajectory that evolves from infinity to $r= 0$ and then returns back to the asymptotic region.  The simplifying assumption $r_\ast = 0$ implies that $\cos\theta = 0$, and constrains the dimensionless ratio
\begin{align}
\eta &= \frac{ P_{\vphi}}{\sqrt{2a^2 E}}~,& \eta^2 &<1~.
\end{align}
For all of these trajectories $\dot \psi = 0$, but since they pass through $r=0$, we obtain 
\begin{align}
(\Delta\psi)_{\infty} = \pi~.
\end{align}
The change in $\vphi$ is more complicated.  After a change of variables $v = \frac{16a^4}{r^4+16a^4}$, we find
\begin{align}
(\Delta\vphi)_{\infty} & = \frac{\eta}2 \int_0^{1} \frac{ dv v^{-3/4} (1-v)^{-1/2}}{\sqrt{1-\eta^2\sqrt{v}}}~.
\end{align}
This is an integral of hypergeometric type, but we will not bother to reduce it further.  It is, of course, simple to
make a series expansion of $(\Delta\vphi)_{\infty}$ for small $\eta$.  The leading term is 
\begin{align}
(\Delta\vphi)_{\infty} & = \frac{\eta}2 \frac{\sqrt{2} \pi^{3/2}}{\Gamma(3/4)^2} + O(\eta^3)~.
\end{align}

\subsection{Multiple geodesics}\label{ss:EHmultiplegeodesics}
An important difference in dynamics on the Eguchi-Hanson space versus the conical geometry is the existence of multiple geodesics connecting a pair of initial and final points.  We have not been able to describe such families of geodesics in complete generality, but, as we now show, it is reasonable to expect that many, perhaps most, points are indeed connected by an infinite number of geodesics.  The geometric feature responsible for this phenomenon is the existence of the trapped geodesics on the exceptional curve:  a family of ``almost trapped'' geodesics can produce arbitrarily large changes of phase in the angular coordinates at the price of an ever increasing geodesic length.

To illustrate this intuition in a precise example, we define a family of geodesics with $P_{\psi} = 0$ and $0<|\eta^2-1| \ll 1$ that start with an initial point $p_0 = (r_0,\theta_0,\phi_0,\psi_0)$ at $t=0$ and end with $p_1 = (r_0,\theta_0,\phi_1,\psi_1)$ at time $t=2T$.  Since $P_{\psi} = 0$, there are just two possibilities for $\psi_1 - \psi_0$:
\begin{align}
\psi_1 -\psi_0 = \begin{cases}  0 & \text{if}~~\eta^2 >1 \\   \pi &\text{if}~~\eta^2 <1 ~. \end{cases}
\end{align}
In the first case $r$ evolves to a turning point $r = r_\ast$ in time $T$; in the second case $r$ reaches the coordinate singularity at $r=0$ at time $T$.  In either case, the time $T$ and the angular shift in coordinates $u=r^2/4a^2$ are given by
\begin{align}
T &= \frac{a}{\sqrt{2E}} \cT(\eta,u_0)~, & 
\cT(\eta,u_0) & = \int_{u_\ast}^{u_0} \frac{du}{\left(\sqrt{1+u^2}-\eta^2\right)^{1/2}}~,&
u_0 & = \frac{r_0^2}{4a^2}~,
\end{align}
and
\begin{align}
\Delta\vphi = \Phi(\eta,u_0) = \eta \int_{u_\ast}^{u_0} \frac{du}{\sqrt{1+u^2}\left(\sqrt{1+u^2}-\eta^2\right)^{1/2}}~,
\end{align}
where
\begin{align}
u_\ast =  \begin{cases}  \sqrt{\eta^4-1} &\text{if}~~ \eta^2 >1 \\ 0 & \text{otherwise}~.\end{cases}
\end{align}
In either case, $|\Phi(\eta,u_0)|$ diverges as $\eta \to 1$, which means we can obtain a sequence of values $\{\eta_1,\eta_2,\eta_3,\ldots\}$, where $\eta_k \to 1$ as $k\to \infty$, such that $2\Phi(\eta_n,u_0) = \phi_1-\phi_0 + 2\pi k$.
This leads to an infinite family of geodesics connecting the two points $p_0$ and $p_1$.  

It is also easy to see that $\cT(\eta_n,u_0)$, and therefore the geodesic length, grows linearly with $k$ for large $k$.  To demonstrate this, we observe that the difference
\begin{align}
\eta \cT(\eta,u_0) - \Phi(\eta,u_0) = \eta  \int_{u_\ast}^{u_0} \frac{du}{\sqrt{1+u^2}}\frac{\sqrt{1+u^2}-1}{(\sqrt{1+u^2}-\eta^2)^{1/2}}~
\end{align}
is finite as $\eta \to 1$.  

With a little more work it is possible to identify the divergence as $\eta^2 \to 1$ more explicitly.\footnote{A straightforward way to obtain this is to examine the $\Phi(\eta,\infty)$ integral; a convenient change of variables turns out to be
$v = (\sqrt{1+u^2}-1)/|1-\eta^2|$.  Making that substitution and then expanding for small $|1-\eta^2|$ leads to this result.}
In either case, we find the leading divergences to be
\begin{align}\label{eq:EHgrowthofgeodesiclength}
\cT &\sim \frac{1}{\sqrt{2}} \log \frac{1}{|1-\eta^2|}~,&
\Phi &\sim \frac{\eta}{\sqrt{2}} \log \frac{1}{|1-\eta^2|}~.
\end{align}

\subsection{Dynamical structure for the Eguchi-Hanson (conifold) in comparison to conic} \label{s:DynStructEH}
This discussion is explicitly for the Eguchi-Hanson geodesic flow.  However, at this level of precision, all of this discussion has obvious and natural parallels in the conifold setting. A reader would not be remiss if she read it again and replaced ``Eguchi-Hanson'' with ``conifold'' after finishing section~\ref{s:conifold}. 

For the Eguchi-Hanson geodesic flow, we have shown that there are three categories of qualitative behaviors: (i) trapped geodesics that are confined to the exceptional curve, (ii) lost geodesics (in either the forward or backwards time), and (iii)  ``standard'' behavior where the r coordinate goes to $\infty$ in both the infinite past and future.  There is no equivalent to the trapped  (fixed $r\neq0$) geodesics  for conics. Case (ii) is a smoothed out version of the radial geodesics that reach the singular point in finite time for the conic.  This introduces new behaviors and individual lost geodesics converge asymptotically to individual trapped geodesics.  For the ``standard'' behavior of (iii), there is a nice parallel with the conic setting: under similar rescaling of the metric (see section~\ref{ss:stabman}), there are boundaries at infinity that naturally identify with the link manifold, $N$ as in section~\ref{ss:boundary}.  However, the connections between the backward and forward boundaries at infinity most likely differ from the conic setting. There, two points on the boundaries (identified with $N\times N$) can be connected by a geodesic orbit in the unit tangent bundle of the cone iff they are distance $\pi$  apart in $N$.  In other words, an infinite future can be, generically, connected to a codimension one family in the infinite past. 

Of these three types of geodesics, geodesics of type (i) are trivial (great circles) and geodesics of type (ii) (a behavior that does not exists for cones) are completely characterized by results in sections \ref{ss:EHgeodesics} and \ref{ss:EHmultiplegeodesics}.  
In particular, any infinite past $\partial^{-\infty} \in N$ (respectively infinite future) on the boundary at infinity can be connected to asymptotic futures (respectively pasts) represented by trapped geodesics simply by the correct choice of momenta ($P_\psi=0$ and $P_\vphi^2=2a^2$ after normalizing energy). In fact, the family of such asymptotic behaviors is precisely the set of geodesics on the exceptional curve (great circles) whose orbit contains $\pi(\partial^{-\infty})$, where $\pi: S^3/\Z_2\rightarrow S^2$ is the projection $(\theta,\vphi,\psi)\to(\theta,\vphi)$.  Note that this is the projection of the Hopf fibration.  
As for type (iii), there are some obvious applications of our computations in  section~\ref{ss:EHmultiplegeodesics} to the connection of asymptotic behaviors for geodesics this type.  However, unlike the case of exact cones, the global picture is still mostly open for the Eguchi-Hanson geometry. Is it possible to connect any infinite past to any infinite future? Is it at least a full dimensional connection?  Or is similarly co-dimensioned like the conic behavior? Is there any non-tautological way of characterizing these connections?

\section{Flows on the resolved conifold} \label{s:conifold}
\subsection{The Candelas-de la Ossa geometry} \label{s:conifoldintro}

In ~\cite{Candelas:1989jb,Candelas:1989js} a smooth Ricci-flat K\"ahler metric is defined on 
the total space of the bundle $\pi:\mathcal{O}(-1)\oplus\mathcal{O}(-1)\to\P^1$.  
This space can be presented as
\begin{align}
M = \frac{ \C^4 \setminus F}{\C^\ast}~,
\end{align}
where the projective coordinates are $[\phi_1,\phi_2,\phi_3,\phi_4]$, the exceptional set is $F =\{\phi_3=\phi_4=0\}$, and the $\C^\ast$ action has charges $Q=(1,1,-1,-1)$. 
If we define affine patches $U$ and $V$ by
\begin{align}
U~,~~ \phi_3\neq 0 &:  \{\phi_1^{(u)} = \phi_3 \phi_1,~\phi_2^{(u)} = \phi_3 \phi_2,~u = \phi_4/\phi_3\}~, \nonumber\\
V~,~~ \phi_4\neq 0 &:  \{\phi_1^{(u)} = \phi_4 \phi_1,~\phi_2^{(u)} = \phi_4 \phi_2,~v = \phi_3/\phi_4\}~,\end{align}
then on the overlap $\C^\ast = U \cap V$, we have
\begin{align}
u &= 1/v~,&
\phi^{(u)}_a & = v \phi^{(v)}_a~~,~~~~a=1,2~.
\end{align}
The space admits a $G=\SU(2)_{\text{fib}}\times\SU(2)_{\text{base}}$ automorphism group, which, in projective coordinates, acts on $(\phi_1,\phi_2)$ as $(\rep{2},\rep{1})$, and on $(\phi_3,\phi_4)$ as $(\rep{1},\rep{2})$.  The metric defined below is a smooth Ricci-flat K\"ahler metric for which these automorphisms are isometries.  

The metric is derived from the $G$-invariant  K\"ahler potential
\begin{align}
\cK^{(u,v)} = f(R) + 4a^2 \log \Lambda^{(u,v)}~,
\end{align}
where $a$ is any constant, $R= (|\phi_1|^2+|\phi_2|^2) (1 + u\ub)$, and $f(R)$ is a function on $M$ such that 
\begin{align}
\gamma^3 + 6 a^2 \gamma^2 = R^2~\label{eq:gammaeq},
\end{align}
with $\gamma = R f'(R)$.
The Calabi-Yau K\"ahler form on $M$ is then
\begin{align}
\omega = i \p\pb f + 4 a^2 \pi^\ast \omega_{\text{FS}}~.
\end{align}
In particular, $\omega^2$ is the globally exact form
\begin{align}
\omega^2 = -\p\pb f \p\pb f + 8ia^2 \p\pb f \pi^\ast \omega_{\text{FS}} = \p\pb \left[-f \p\pb f + 8ia^2 f \pi^\ast \omega_{\text{FS}} \right]~.
\end{align}

We transform the three complex coordinates $u$, $\phi_1$, and $\phi_2$ into a new real spherical coordinate system of $r,\theta_1,\theta_2,\vphi_1,\vphi_2$, and $\psi$ as follows.  Let
\begin{align}
u &=  \frac{\cos\theta_1-1}{\sin\theta_1} e^{-i\vphi_1}~,
\end{align}
with $0\leq \theta_1\leq\pi$ and $0\leq\vphi_1<2\pi$. Let 
\begin{align}
 \phi_1 & = \sqrt{\frac{R(1+\cos\theta_1)}{2}}  \cos(\theta_2/2) e^{\frac{i}2(\psi+\vphi_1+\vphi_2)}~,&\phi_2 & =\sqrt{\frac{R(1+\cos\theta_1)}{2}} \sin(\theta_2/2) e^{\frac{i}2(\psi+\vphi_1-\vphi_2)}~,
\end{align}
where $0\leq\theta_2<\pi$, $0\leq\vphi_a<2\pi$ and $0\leq\psi<4\pi$, are angular coordinates, and $R =\left(\frac{2}{3}\right)^{3/2} r^3$ define the new radial coordinate $r$.  Let $p_r,p_{\theta_1},p_{\theta_2},p_{\vphi_1},p_{\vphi_2}$ and $p_\psi$ denote the corresponding conjugate momenta.

If we let $y = \gamma/R^{2/3}$ and $\ep = 6 a^2/R^{2/3}$, then (\ref{eq:gammaeq}) becomes the dimensionless equation
\begin{align}
y^3 + \ep y^2 = 1~.
\end{align}
Note that $y\in[0,1]$ is an increasing function for $r\in[0,\infty)$.  It will often be convenient to compute with respect to this `normalized' radial coordinate.

In terms of our new $r$ and our chosen angular coordinates, the metric takes the form

\begin{align}
g &= \frac{1}{A_1A_2} dr^2 + \frac{A_1 r^2}{6} (d\theta_1^2+\sin^2\theta_1 d\vphi_1^2) + \frac{A_2r^2}{6}(d\theta_2^2+\sin^2\theta_2 d\vphi_2^2) + \frac{r^2}{9A_1A_2} \Psi^2~,
\end{align}
where
\begin{align}
\Psi & =d\psi+\cos\theta_1d\vphi_1+\cos\theta_2d\vphi_2~, &&A_1  = y+\frac{2}{3}\ep~,&&
A_2 = y~,&
\ep = \frac{9a^2}{r^2}~.
\end{align}

For large $r$, the metric is asymptotic to the conical metric obtained by setting $a=0$.  In particular, both $A_1$ and $A_2$ asymptote to $1$, and the metric is asymptotic to
\begin{align}
g &= dr^2 + r^2\left(\frac{1}{6} (d\theta_1^2+\sin^2\theta_1 d\vphi_1^2) + \frac{1}{6}(d\theta_2^2+\sin^2\theta_2 d\vphi_2^2) + \frac{1}{9} \Psi^2\right)~,
\end{align}
the warped product on $\R_{\ge 0} \times N$ from (\ref{e:alemetric}).  It is this (singular) space with this metric that we refer to as `the conifold', with the smooth manifold $M$ as its small resolution.  In fact, the metric on the link $N$ is a round metric on both copies of $S^2$ together with $\Psi^2$ which is the product metric on a circle fibration.\footnote{The scaling factors $\frac16,\frac16$ and $\frac19$ make the metric on $N$ Einstein, a necessary and sufficient condition for the metric on $M$ to be Ricci-flat. }  In fact, the link $N$ is the well known space $T^{1,1}$, so named after the pair of relatively prime coefficients of $\cos\theta_ad\vphi_a$ in $\Psi$.  The  $S^1$  from the circle bundle is fibered diagonally  over $S^2\times S^2$ as a Hopf fibration making $N$ topologically, but not geometrically, $S^3\times S^2$.  For more details on the definition of the conifold, its link $N$, and its  metric see \cite{Candelas:1989js}.


Once again, this metric has a coordinate singularity at $r=0$, and we can choose new coordinates near that singularity.  Let 
\begin{align}
x_1+i y_1 & = z_1  =\sqrt{\frac{r^3}{27a}} \cos\ff{\theta_2}{2} e^{\frac{i}2(\psi+\vphi_2)}~, &
x_2 + i y_2 & = z_2  = \sqrt{\frac{r^3}{27a}}\sin\ff{\theta_2}{2} e^{\frac{i}2(\psi-\vphi_2)}~.
\end{align}
For small values of $r$
\begin{align}
g &=  a^2 (d\theta_1^2 + \sin^2\theta_1 d\vphi_1^2) 
\nonumber\\
&+6\left[dx_1^2 + dy_1^2 +dx_2^2 + dy_2^2 + \cos\theta_1 d\vphi_1 ( x_1 dy_1 - y_1 dx_1 +x_2 dy_2 - y_2 dx_2)\right]+\nonumber\\
&+O (\sum x_a^2+y_a^2)~,
\end{align}
which is smooth and reduces appropriately on the exceptional curve.

The inverse metric is found to be
\begin{align}
g^{-1} & = A_1 A_2 \left(\frac{\p}{\p r}\right)^2 
+ \frac{6}{r^2 A_1} \left(\frac{\p}{\p \theta_1}\right)^2
+ \frac{6}{r^2 A_2} \left(\frac{\p}{\p \theta_2}\right)^2
+\frac{6}{r^2 A_1 \sin^2\theta_1} \Upsilon_1^2
+\frac{6}{r^2 A_2 \sin^2\theta_2} \Upsilon_2^2 \nonumber\\
&\qquad + \frac{9 A_1 A_2 }{r^2}  \left(\frac{\p}{\p \psi}\right)^2,
\end{align}
where
\begin{align}
\Upsilon_a = \frac{\p}{\p \vphi_a} - \cos\theta_a\frac{\p}{\p\psi}~.
\end{align}

Lastly, we introduce the 
Killing vector fields for the conifold.  Since the metric is independent of the coordinates $\varphi_a$ and $\psi$, $V_\psi  = \frac{\p}{\p\psi}$ and the pair $V_{3a}=\frac{\p}{\p\vphi_a}$ are Killing fields.  In addition, the following four fields are Killing:
\begin{align}
V_{1a}&=-\sin\vphi_a\frac\p{\p\theta_a}-\cot\theta_a\cos\vphi_a\frac\p{\p\vphi_a}+\frac{\cos\vphi_a}{\sin\theta_a}\frac\p{\p\psi}~,\nonumber\\
V_{2a}&=\cos\vphi_a\frac\p{\p\theta_a}-\cot\theta_a\sin\vphi_a\frac\p{\p\vphi_a}+\frac{\sin\vphi_a}{\sin\theta_a}\frac\p{\p\psi}~.
\end{align}
These fields are found by perturbing Killing fields on $S^2$ to also annihilate $\Psi$.  The $V_{1a}, V_{2a}$ and $V_{3a}$ each satisfy the $SU(2)$ algebra relations, giving an overall $SU(2)\times SU(2)$ symmetry.

\subsection{Hamiltonian and Conserved charges}\label{ss:conifoldHamiltonian}

We read off the Hamiltonian from the form of the inverse metric:
\begin{align}
2H &= A_1A_2 p_r^2+\frac{6}{r^2 A_1} p_{\theta_1}^2+\frac{6}{r^2 A_2} p_{\theta_2}^2 
+\frac{6}{r^2 A_1\sin^2\theta_1}(p_{\vphi_1}- \cos\theta_1 p_{\psi})^2 
\nonumber\\
&\qquad
+\frac{6}{r^2 A_2 \sin^2\theta_2}(p_{\vphi_2}-\cos\theta_2 p_{\psi})^2
+ \frac{9A_1A_2}{r^2}  p_{\psi}^2~.
\end{align}
We observe that it is independent of $\psi$ and both of the $\vphi_a$, and the Killing fields of the metric
yield seven conserved quantities:
\begin{align}
P_\psi  &=p_\psi= \frac{A_\psi r^2}{9} \left(\dot\psi+\cos\theta_1\dot\vphi_1+\cos\theta_2\dot\vphi_2\right)~, \nonumber\\
P_{\vphi_a} & =p_{\vphi_a}= \frac{A_a r^2}{6}  \sin^2\theta_a \dot\vphi_a+P_\psi \cos\theta_a~,\nonumber\\
P_{1a} & = -\frac{A_a r^2}{6}(\sin\vphi_a\dot\theta_a +\cos\theta_a\sin\theta_a \cos\vphi_a \dot\vphi_a) + P_\psi \cos\vphi_a\sin\theta_a~,\nonumber\\
P_{2a} & = \frac{A_a r^2}{6} (\cos\vphi_a\dot\theta_a -\cos\theta_a\sin\theta_a \sin\vphi_a\dot\vphi_a) + P_\psi\sin\vphi_1\sin\theta_a ~.
\end{align}
These satisfy for $a=1,2$
\begin{align}
(P_{1a}\sin\vphi_a +P_{2a}\cos\vphi_a) \sin\theta_a + P_{\vphi_a} \cos\theta_a = P_\psi~,
\end{align}
which takes an elegant form in terms of
\begin{align}
\bX_a & = \begin{pmatrix} \sin\theta_a \cos\vphi_a \\ \sin\theta_a \sin\vphi_a \\ \cos\theta_a \end{pmatrix}~, &
\bP_a & = \begin{pmatrix} P_{1a} \\ P_{2a} \\ P_{\vphi_a} \end{pmatrix}~.
\end{align}
We now see that the coordinates and conserved momenta
satisfy the $\SO(3)\times SO(3)$--invariant constraints
\begin{align}
\bX_a \cdot \bP_a & = P_{\psi}~.
\end{align}
We can see that each value of $P_\psi$ corresponds to a (not necessarily great) circle on each of the $2$-spheres.  This is just a generalization of the structure we already observed for the Eguchi-Hanson geometry, and, as in that case, we can use the conserved quantities to solve for the dynamics of the angular coordinates. 
Let 
\begin{align}
P_{1a} & = S_a \cos\alpha_a ~,&
P_{2a} & = S_a \sin\alpha_a~,
\end{align}
so that $p_{\theta_a}$ is determined by
\begin{align}
p_{\theta_a} & = S_a \cos(\vphi_a + \alpha_a)~.
\end{align}
Note that  $P_{1a}$ and $P_{2a}$ conservation implies that the $\vphi_a(t)$ are determined by the $\theta_a(t)$ and conserved quantities.  In particular, the angles $\alpha_a$ can be thought of as the initial angles $\vphi_a(0)$. 

The total energy can now be rewritten as
\begin{align}
2E &= A_1A_2 p_{r}^2+ \frac{9A_1 A_2}{r^2} P_{\psi}^2+ \frac{6(S_1^2 + P_{\vphi_1}^2 - P_{\psi}^2)}{r^2 A_1} +  \frac{6(S_2^2 + P_{\vphi_2}^2 - P_{\psi}^2)}{r^2 A_2}  \nonumber\\
& =  A_1A_2 p_{r}^2 + \frac{6(S_1^2 + P_{\vphi_1}^2)}{r^2 A_1}  + \frac{6(S_2^2 + P_{\vphi_2}^2)}{r^2 A_2}  
+ \left( \frac{9A_1 A_2}{r^2} - \frac{6}{r^2 A_1} - \frac{6}{r^2 A_2} \right) P_{\psi}^2~.
\end{align}
As energy is a conserved quantity, this implies that $p_r$ is determined as a function of $r$ and the integration constants.  


\subsection{Properties of geodesics}

Employing the $\SO(3)$ invariance of the momenta noted in the previous section, we take $\bP_a$ to lie along the $z$--axis.  As a consequence, $P_{1a} = P_{2a} = 0$, which implies that $p_{\theta_a} = 0$ and therefore the $\theta_a$ are constants. 
Moreover, $P_{\vphi_a} \cos\theta_a = P_{\psi}$.  Applying this to the  $\vphi_a$ equations of motion determines the rates of precession of the $\vphi_a$s:
\begin{align}
\label{eq:phidot}
\dot \vphi_a &= \frac{\p H}{\p p_{\vphi_a}}= \frac{6}{r^2 A_1 \sin^2\theta_a} (P_{\vphi_a}-\cos\theta_a P_{\psi}) = \frac{6 P_{\vphi_a}}{r^2 A_a}~.
\end{align}
Similarly, the rate of procession of $\psi$ is 
\begin{align}
\label{eq:psidot}
\dot\psi & = \frac{\p H}{\p p_{\psi}} = \frac{9 A_1 A_2}{r^2} P_{\psi}~,
\end{align}
which is a function of only $r$, so determines $\psi$ in terms of its value at $t=0$.

After employing all of these reductions, the evolution of the radial coordinate $r$ is governed by the effective Hamiltonian
\begin{align}
\label{eq:effectiveHamiltonian}
H_{\text{eff}} & =  \frac12A_1A_2 p_{r}^2 + U_{\text{eff}}~,
\end{align}
where the effective potential $U_{\text{eff}}$ is the following function of $r$, conserved quantities, $\cos\theta_1$ and $\cos\theta_2$:
\begin{align}\label{eq:effectivePotential}
U_{\text{eff}}&= \frac{3\sin^2\theta_1P_{\vphi_1}^2}{r^2 A_1}  + \frac{3\sin^2\theta_2P_{\vphi_2}^2}{r^2 A_2}  
+ \left( \frac{9A_1 A_2}{2r^2}  \right) P_{\psi}^2~.
\end{align}
This implies
\begin{align}
\label{eq:rdot}
\dot r = \frac{\p H_{\text{eff}}}{\p p_{r}} = A_1 A_2 p_{r} = A_1A_2 P_r(r)~,
\end{align}
and conservation of energy determines 
\begin{align}
p_r&=\pm\sqrt{\frac{2(E-U_{\text{eff}})}{A_1A_2}}~.
\end{align}

\begin{figure}
\begin{center}
\includegraphics[width=7cm]{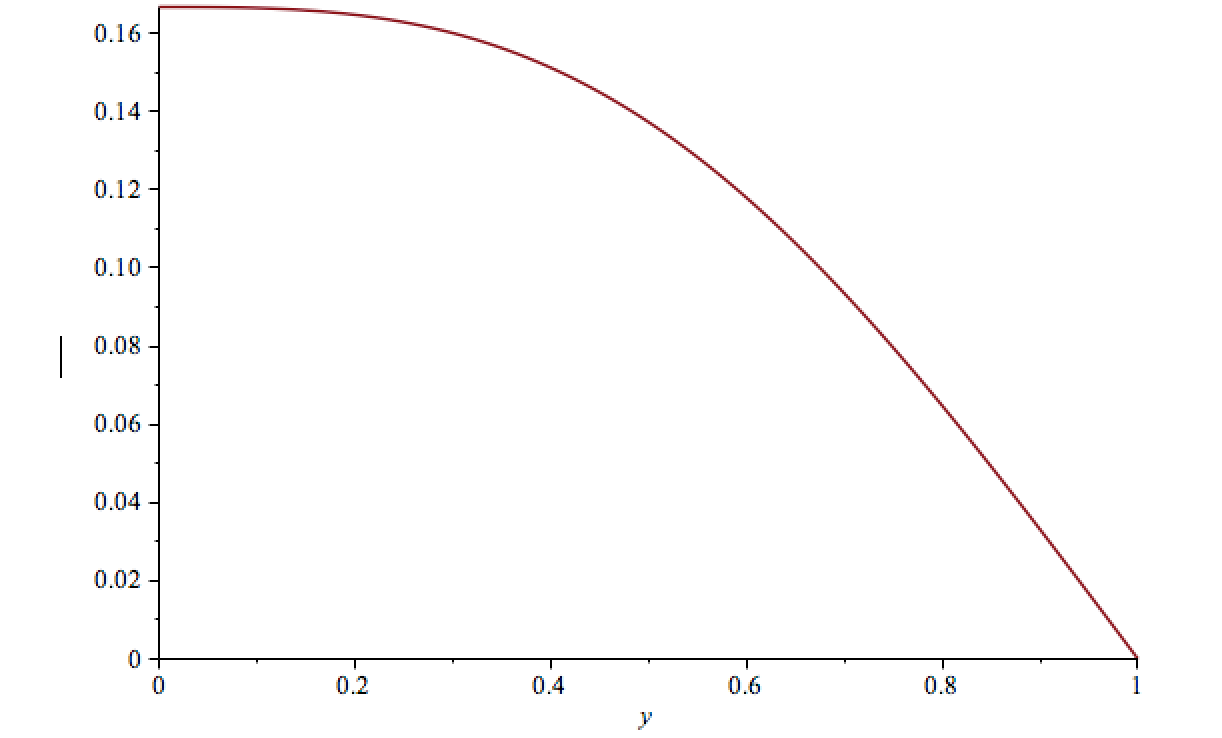}  
\includegraphics[width=7cm]{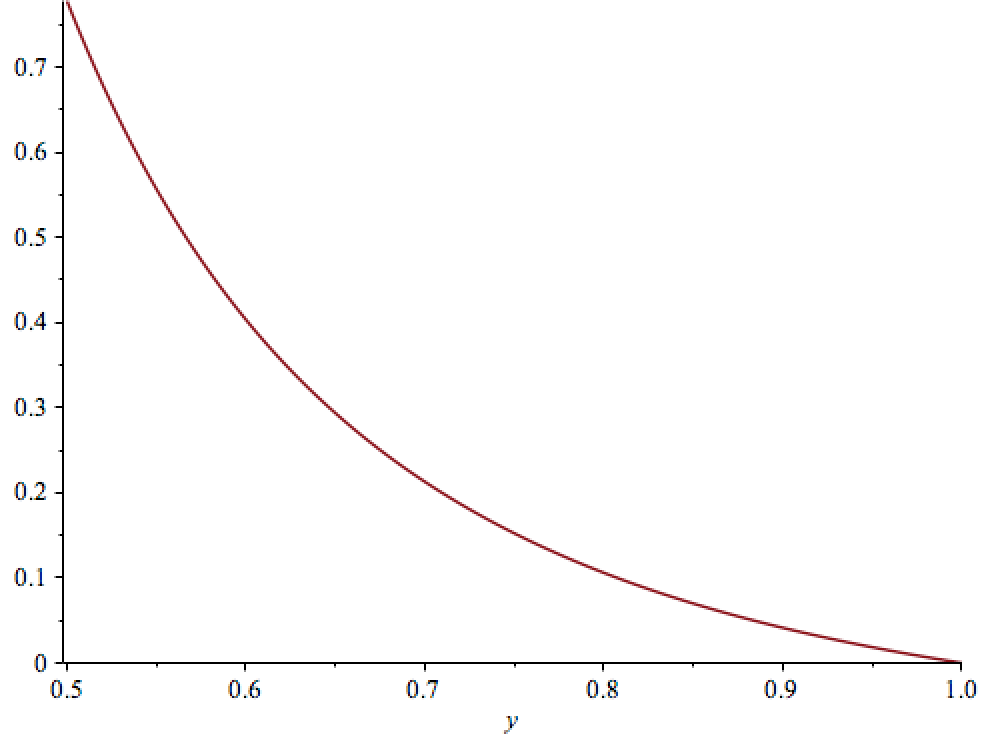}
\end{center}
\caption{A plot of $a^2\cU_1$ as a function of $y$ (left) and $a^2\cU_2$ (right).  Note that $\cU_2$ blows up to $+\infty$ as $y\to 0$. }\label{f:U1U2}
\end{figure}

\begin{figure}
\begin{center}
\includegraphics[width=8cm]{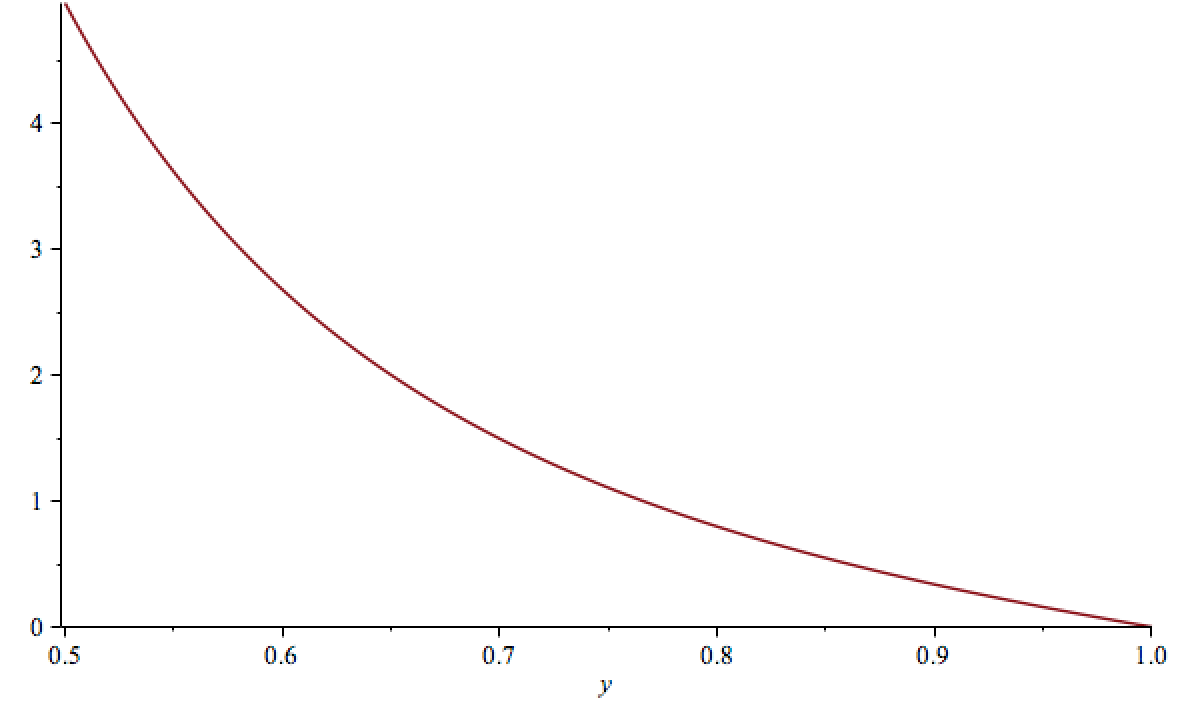}
\end{center}
\caption{A plot of $a^2\cU_3$ as a function of $y$.}\label{f:U3}
\end{figure}

After a change variables ($r$ to $y$),  we are left with our final expression for the total energy 
\begin{align}\label{eq:totalenergy}
2a^2H&=\frac{ 2 (1-y^3)^3}{27 y(y^3+2)}p_y^2+\frac{1-y^3}{y^3+2}(P_{\vphi_1}^2-P_\psi^2)+\frac{1-y^3}{3y^3}(P_{\vphi_2}^2-P_\psi^2)+\frac{(1-y^3)(y^3+2)}{6 y^3}P_\psi^2~,
\end{align}
where $P_\psi=\cos\theta_aP_{\vphi_a}$.  Let
\begin{align}\label{eq:FandU}
\cF & = \frac{ 4 (1-y^3)^3}{27 y(y^3+2)}~,&
\cU_1 & = \frac{1-y^3}{y^3+2}~,&
\cU_2 & = \frac{1-y^3}{3y^3}~,&
\cU_3 & = \frac{(1-y^3)(y^3+2)}{6 y^3}~.
\end{align}The graphs of $\cU_1$, $\cU_2$ and $\cU_3$ are given in figures \ref{f:U1U2} and \ref{f:U3}.  Notice that $\cU_2$ and $\cU_3$ diverge at $y\to 0$, and none of the functions have critical points for $y\neq 0,1$.  

Since energy is conserved and $y$ is defined on $[0,1]$, all of these coefficients are positive and there are no orbits with bounded radius (other than, possibly, $r=0$).  The qualitative dynamics for $r$ is very simple:  if $\sin^2\theta_2P_{\vphi_2}$ or $P_{\psi}\neq 0$ there is a unique turning point with $r_*>0$.  If both $\sin^2\theta_2P_{\vphi_2}=0$ and $P_{\psi}=0$, then for sufficiently large $E$, there will be orbits that can reach $r=0$.\footnote{  These conclusions follow from the $\cU_1$, $\cU_2$, and $\cU_3$ given in the figures.}  
Understanding these ``lost orbits" requires us to think more carefully about the $r=0$ coordinate singularity.

Towards that end, 
the equation of motion for $y$ is now
\begin{align}
\label{eq:ydot}
\dot y & = \frac{\cF}{a^2} p_y~.
\end{align}
Since $\cF \ge 0$ for $0\le y \le 1$, it follows that if $p_y <0$ at $t=0$, then $y$ will decrease until it reaches a turning value $y_\ast$ (for which $p_y  = 0$).  Such a turning point is determined by a solution to
\begin{align}
\label{eq:turningpoints}
1 & =  \frac{\sin^2\theta_1P_{\vphi_1}^2}{E} \cU_1(y)  + \frac{\sin^2\theta_2P_{\vphi_2}^2}{E} \cU_2(y) + P_\psi^2 \cU_3(y)~.
\end{align}
Letting $L$ be the right hand side of (\ref{eq:turningpoints}), notice that $L$ is $0$ at $y=1$, and decreasing on $[0,1]$.  
Since $\cU_{1,2,3}$ are all non-negative,  and $\cU_{2,3}$ blow up as $y^{-3}$ for small $y$, this turning point will be unique and will have $y_\ast >0$ if $\sin^2\theta_2P_{\vphi_2}^2>0$ or $P_\psi^2 >0$.   As long as this is the case, the turning point is reached in finite time, and the trajectory will then evolve to $y=1$ as $t\to\infty$.

Suppose $\sin^2\theta_2P_{\vphi_2}=P_\psi= 0$.  Since $\cU_1(0) = 1/2$, there are three possible cases.  If $\sin^2\theta_1P_{\vphi_1}^2 > 2$, then $L(0)>1$ and the results are the same:  there is a turning point with $y_\ast > 0$.  The other two cases, $\sin^2\theta_1P_{\vphi_1}^2=2$ and $\sin^2\theta_1P_{\vphi_1}^2<2$, will be dealt with in subsequent subsections.

\subsubsection*{Near the exceptional divisor}

We need a  change of coordinates in order to carefully examine the dynamics at $r=0$.
Letting
\begin{align}
r = 3 a^{1/3} \rho^{2/3}~ \implies \ep = \left(\frac{a}{\rho}\right)^{4/3} = \frac{1}{\ept^2} ~,\qquad \ept = \left(\frac{\rho}{a}\right)^{2/3}~,
\end{align}
we rewrite the defining $y$ equation in terms of $y = \ept w$ to find that $w$ satisfies
\begin{align}
w^2 + (\ept)^3w^3 & = 1~ \implies w = 1 -\ff{1}{2} \ept^{\,3} + O(\ept^6)~.
\end{align}
We then define rectangular coordinates for $\rho, \theta_2,\vphi_2$, and $\psi$ as follows:
\begin{align}
x_1+i y_1 & = z_1  =\rho \cos\ff{\theta_2}{2} e^{i\left(\frac{\psi+\vphi_2}2\right)}~, &
x_2 + i y_2 & = z_2  = \rho\sin\ff{\theta_2}{2} e^{i\left(\frac{\psi-\vphi_2}2\right)}~,
\end{align}
(recall that $0 \le \theta_2 \le \pi, ~0 \le \psi  <4\pi$, and $0 \le \vphi_2 < 2\pi$).

The full Hamiltonian is now
\begin{align}
2H & = \frac{1}{a^2 \At_1} \left[ p_{\theta_1}^2 + \frac{p_{\vphi_1} - 2\cos\theta_1 p_\psi ^2}{\sin^2\theta_1}  \right] \nonumber\\
&\qquad +\frac{1}{6 \At_2} \left(p_{x_1}^2+p_{y_1}^2 + p_{x_2}^2 + p_{y_2}^2 \right)\nonumber\\
&\qquad +\frac{\At_3}{6}\left(\rho^2 p_\rho^2 +4p_\psi^2\right)~,
\end{align}
where
\begin{align}
\At_1(w) = 1 + \frac{3}{2} \ept^{\,3} w~,&&\At_2(w) = w~.
\end{align}

Recalling the case of Eguchi-Hanson in section \ref{ss:EHgeodesics}, we see that once again

\begin{enumerate}

\item There are trapped geodesics.  This formulation of the Hamiltonian makes it clear that  $x_a = y_z = 0$ and $p_{x_a} = p_{y_a} = 0$ are preserved by the flow.  
In particular, noting that
\begin{align}
\At_1 & = 1+\frac{3}{2} \left(\frac{\rho}{a}\right)^2 + O(\rho^4)~, &
\At_2 & = 1-\frac{1}{2} \left(\frac{\rho}{a}\right)^2 + O(\rho^4)~.
\end{align}
means that everything is well defined at $\rho=0$.

\item Trajectories headed towards 
$\rho = 0$ with $\dot \rho < 0$ pass through the coordinate singularity.  These are trajectories where $p_{\vphi_2} = p_{\psi} = p_{\theta_2} = 0$, which implies that they pass through the origin in $\R^4$ with $\vphi_2,\psi,\theta_2$ constant
 and re--emerge with $\dot\rho>0$ and the angle $\psi = \psi+2\pi$.  The latter just accomplishes $(x_1,y_1,x_2,y_2) \to (-x_1,-y_1,-x_2,-y_2)$ in rectangular coordinates. 
 \end{enumerate}

\subsubsection*{Lost geodesics: $P_\psi=0, P_{\vphi_2}=0, E=P_{\vphi_1}$}

If $E=U_{\text{eff}}$, the trajectory approaches $y=0$ in infinite time.
Expanding (\ref{eq:FandU}) for small $y$ results in
\begin{align}
\cF &= \frac{2}{27y} + O (y^2)~, & 2\cU_1 & = 1 - \frac{3y^3}{2} +O(y^6)~,
\end{align}
so that energy conservation implies that 
\begin{align}
p_y^2 & = \frac{81y^4 a^2 E}{2} + O (y^6)~.
\end{align}
Therefore
\begin{align}
p_y = - 9y^2 a\sqrt{\frac{E}{2}} + O(y^3)~,
\end{align}
and  the $y$ equation of motion is $\dot y = - k y + O(y^5)$ for $k$ a positive constant.  In other words, $y \approx y_0 e^{-kt}$ and the trajectory only reaches $y=0$ in infinite time.

\subsubsection*{Radial geodesics: $P_{\phi_a}=0$}

Using the $\SO(3)\times\SO(3)$ invariance from section \ref{ss:conifoldHamiltonian}, we can choose coordinates such that $P_{1a}=P_{2a}=S_a=0$ and $P_{\vphi_a}\cos\theta_a=P_\psi$.    If, in addition, we assume that $P_\psi=0$, then (just like it did with Eguchi-Hanson), the entire effective potential  vanishes (\ref{eq:effectivePotential}), leaving us with 
\begin{align}
\dot r=A_1A_2p_r~,&&p_r=\pm\sqrt{\frac{2E}{A_1A_2}}~.
\end{align}
The geodesics with $r\neq0$ and $p_r\geq0$ are only interesting in the limit (the manifold at infinity, discussed in sections~\ref{ss:boundary} and~\ref{s:DynStructEH}), so here, we assume $p_r=-\sqrt{\frac{2E}{A_1A_2}}$. 

In this case, $\sin^2\theta_1P_{\vphi_1}=\sin^2\theta_2P_{\vphi_2}^2=P_\psi^2=0$, so (\ref{eq:turningpoints}) has no solutions.  Therefore, $p_r\neq0$ always, and there are no turning points.  
Since geodesics have constant speed in the metric, these orbits take a finite amount of time to reach $r=0$.
In particular, if we let $r(0)=r_0$ and $p_r(0)=-\sqrt{2E/A_1(r_0)A_2(r_0)}$, then the trajectory will reach $r=0$ at a time $T$ given by 
\begin{align}
T&=\int_{r_0}^0\frac{dr}{\dot r}=\int_0^{r_0}\frac{dr}{\sqrt{2EA_1A_2}}.
\end{align}
In terms of the normalized radius $y$, time is given by 
\begin{align}
\sqrt{2E}~aT=\int_0^{y_0}\frac{dy}{\sqrt{\cF}}&=\int_0^{y_0}\sqrt{\frac{27 y(y^3+2)}{ 2 (1-y^3)^3}}dy~,
\end{align}
which implies that $T$ is finite as long as $y_0<1$ (that is, for any trajectory that doesn't start at the manifold at infinity).  We can also conclude that the trajectory evolves as
\begin{align}
\underbrace{(r_0,{\theta_1}_0,{\theta_2}_0,{\vphi_1}_0,{\vphi_2}_0,\psi_0)}_{t=0}  \to  \underbrace{(r_0,{\theta_1}_0,{\theta_2}_0,{\vphi_1}_0,{\vphi_2}_0,\psi_0+2\pi)}_{t=2T}~.
\end{align}

\subsubsection*{Angular shifts} 

For a general geodesic, we can not assume that $P_\psi=0$ (and consequently, that $\vphi_1,\vphi_2,\psi$ remain constant).  In particular, there are non-trivial angular shifts.  Consider a path starting at $r(0)=r_0$ with $p_r<0$.  Unless the energy is exactly correct ($E=U_{\text{eff}}$), this trajectory reaches a turning point (either $r_*\neq0$ or $r_*=0$) in some finite time $T$.  

Just like Eguchi-Hanson, it is again possible to set up integrals for the time $T$, and, just like it was in the previous section, it is more enlightening to work with  $y$  than $r$:
\begin{align}
T&=\int_{y_0}^{y_*}\frac{dy}{\dot y}=-\int_{y_0}^{y_*}\frac{dy}{\sqrt{2\cF(E-U_{\text{eff}})}}=\int_{y_*}^{y_0}\frac1{\sqrt{2(E-U_{\text{eff}})}}\frac{dy}{\sqrt{\cF}}~.
\end{align}

Using (\ref{eq:phidot}), we compute 
\begin{align}
\Delta\vphi_1&=\vphi_1(T)-\vphi_1(0)=\int_0^T\frac{6P_{\vphi_1}dt}{r^2A_1}=6P_{\phi_1}\int_{y_*}^{y_0}\frac{1-y^3}{3a^2(y^3+2)}\frac{dy}{\sqrt{2\cF(E-U_{\text{eff}})}}\\
&=6P_{\vphi_1}\int_{y_*}^{y_0}\frac{1-y^3}{3a^2(y^3+2)}\frac{dy}{\sqrt{2(E-U_{\text{eff}})}}\sqrt{\frac{27 y(y^3+2)}{ 2 (1-y^3)^3}}~,
\end{align}
which converges if $E\neq U_{\text{eff}}$.

On the other hand, 
\begin{align}
\Delta\vphi_2=\int_0^T\frac{6P_{\vphi_2}dt}{r^2A_2}=6P_{\phi_2}\int_{y_*}^{y_0}\frac{1-y^3}{9a^2y^3}\frac{dy}{\sqrt{2\cF(E-U_{\text{eff}})}}~,
\end{align}
which is only finite if both $y_*> 0$ and the not-a-lost-orbit condition holds.  However, this is not a problem, as we note that $y_*=0$ implies $P_\psi=\sin\theta_2P_{\vphi_2}=0$ and, by choosing our coordinates appropriately, $P_{\vphi_2}\cos\theta_2=0$.  That is, $y_*=0$ implies that $P_{\vphi_2}=0$ and therefore, $\Delta\vphi_2=0$.

Similarly, using (\ref{eq:psidot}), we have
\begin{align}
\Delta\psi&=\int_0^T\frac{9 A_1 A_2dt}{r^2} P_{\psi}=P_\psi\int_{y_*}^{y_0}\frac{(1-y^3)(y^3+2)}{3a^2y^3}\frac{dy}{\sqrt{2\cF(E-U_{\text{eff}})}}~,
\end{align}
which converges if $y_*>0$ and is zero if $y_*=0$.

\subsection{Multiple geodesics}

Just like  the Eguchi-Hanson space, we find that the conifold  possesses multiple, arbitrarily long geodesics connecting many pairs of initial and final points.  Although we have been unable to determine the dynamics of such orbits completely, again, we expect that many, perhaps most, points are connected by an infinite number of geodesics.  To locate geodesics of this type, we define families of curves that approach a trapped geodesic, but instead of living entirely on the exceptional curve, these families either pass through $r=0$ or turn around at some small $r_*$ while still imitating the dynamics of the trapped curves.  This allows us to produce arbitrarily large changes in phase in either of the $\theta_1$ or  $\vphi_1$ angular coordinates, which results in arbitrarily long geodesics. 

To simplify computations, we assume that the angular coordinates that vanish on the exceptional curve are constant (namely, $P_{\theta_2}=P_{\vphi_2}=P_\psi=0$).  This, in turn, implies that we can choose coordinates such that $\cos\theta_1P_{\vphi_1}=0$, so either $P_{\vphi_1}=0$ or $\theta_1=\frac\pi2$.  We assume the latter, which makes the effective potential (\ref{eq:effectivePotential}) and the total energy (\ref{eq:totalenergy}) 
\begin{align}
U_{\text{eff}}&=\frac{1-y^3}{a^2(y^3+2)}P_{\vphi_1}^2~,&&
E=\frac{ 2 (1-y^3)^3}{27 a^2y(y^3+2)}p_y^2+\frac{1-y^3}{a^2(y^3+2)}P_{\vphi_1}^2~.
\end{align}
Notice that $U_{\text{eff}}(0)=\frac{P_{\vphi_1}^2}{2a^2}$, so such a trajectory has a turning point if $E<\frac{P_{\vphi_1}^2}{2a^2}$, is a lost orbit if $E=\frac{P_{\vphi_1}^2}{2a^2}$, and passes through the singularity if $E>\frac{P_{\vphi_1}^2}{2a^2}$.
Let 
\begin{align}
\eta_1&=\frac{P_{\vphi_1}}{\sqrt{2a^2E}}~
\end{align}
for convenience.  Notice this is exactly the same constant with exactly the same relevance as $\eta$ for Eguchi-Hanson.
Also for convenience, we work with dimensionless time $\cT=T/\sqrt{2E}a$, so the time integral becomes
\begin{align}
\cT&=\int_{y_*}^{y_0}\frac{dy}{\sqrt{a^2\cF(1-\cU_{\text{eff}}/E)}}~.
\end{align}

We construct the family of geodesics analogous to the one we constructed for Eguchi-Hanson (section \ref{ss:EHmultiplegeodesics}).  Assume that $0<|\eta_1^2-1| \ll 1$, our initial point is \\$p_0=(y_0,{\theta_1}_0,{\theta_2}_0,{\vphi_1}_0,{\vphi_2}_0,\psi_0)$ at time $t=0$, and our end point is $p_1=(y_0,{\theta_1}_0,{\theta_2}_0,{\vphi_1}_1,{\vphi_2}_0,\psi_1)$ at time $t=2T$.  (Recall that we are choosing coordinates such that ${\theta_1}_0=\frac\pi2$.)  Again, since $P_\psi=0$, there are two possibilities for the terminal value $\psi_1$, which depend on whether our geodesic passes through the singularity or has a turning point with $r_*>0$.  In particular,
\begin{align}
\psi_1 -\psi_0 = \begin{cases}  0 & \text{if}~~\eta_1^2 >1 \\   2\pi &\text{if}~~\eta_1^2 <1 ~. \end{cases}
\end{align}

Let 
\begin{align}
\Delta\vphi_1&=\Phi(\eta_1,y_0)=\int_{y_*}^{y_0}\frac{\dot\vphi_1}{\dot y}dy=\eta_1\sqrt{\frac{27}{1+2\eta_1^2}}\int_{y_*}^{y_0}\sqrt{\frac{y}{(1-y^3)(y^3+\varepsilon)}}dy~,
\end{align}
where $\varepsilon=2\left(\frac{1-\eta_1^2}{1+2\eta_1^2}\right)$.  Since $y_*$ is the largest root of $y^3+2-2\eta_1^2(1-y^3)$, notice that
\begin{align}
y_*^3 = \begin{cases}  -\varepsilon & \text{if}~~\eta_1^2 >1 ~~(\varepsilon<0)\\   0 &\text{if}~~\eta_1^2 <1 ~~(\varepsilon>0)~. \end{cases}
\end{align}
In either case, $|\Phi(\eta_1,y_0)|$ diverges at $y_*$ as $\eta_1\to 1$, so (by continuity) we can find a sequence of values $\{{\eta_1}_1,{\eta_1}_2,{\eta_1}_3,...\},$ converging to $1$, such that $2\Phi(\eta_k,y_0)={\vphi_1}_1-{\vphi_1}_0+2\pi k.$  These values for $\eta_1$ define our infinite family of geodesics connecting $p_0$ to $p_1$.

In fact, dimensionless time (and therefore geodesic length) grow linearly in $k$ for large $k$. To see this, first notice that $\cT$ and $\Phi$ grow at the same rate since
\begin{align}
\eta_1\cT-\Phi(\eta_1,y_0)=\frac{\sqrt{27}}2\eta_1\int_{y_*}^{y_0}\frac{3y^{\frac32}dy}{(1-y^3)^{\frac32}(y^3+2-2\eta_1^2(1-y^3))^\frac12}~,
\end{align}
converges as $\eta_1\to 1$, and therefore, the two quantities differ by a constant.  
To measure their common rate of growth, we work with the simpler of the two quantities and compute $\Phi(\eta_1,1)$ (the shift in $\vphi_1$ for an orbit starting at the manifold at infinity and ending at its turning point). 
We find that the leading divergences of $\cT$ and $\Phi$ as $\eta_1^2\to 1$ to be
\begin{align}
\cT&=\log\frac1{|1-\eta_1^2|},&&\Phi\sim\eta_1\log\frac1{|1-\eta_1^2|}~,
\end{align}
so linear growth.  Notice that this growth rate only differs from the growth of geodesic length for the corresponding Eguchi-Hanson space example (\ref{eq:EHgrowthofgeodesiclength}) by a factor of $1/\sqrt{2}$.  


Thus, as promised in section~\ref{s:DynStructEH}, at this level of analysis we find the same qualitative conclusions for the asymptotic dynamics of the resolved conifold as for those of the Eguchi-Hanson space.  It will be interesting in future work to determine how far the similarities persist, and whether there are more subtle aspects of the asymptotic dynamics in which the two geometries differ.


\begin{thebibliography}{10}

\bibitem{Candelas:1989jb}
P.~Candelas, P.~S. Green, and T.~Hubsch, ``Connected {C}alabi-{Y}au
  compactifications (other worlds are just around the corner),'' in {\em
  {Strings 88: A Superstring Workshop College Park, Maryland, May 24-28,
  1988}}, pp.~0155--190.
\newblock
1989.
\newblock

\bibitem{Candelas:1989js}
P.~Candelas and X.~C. de~la Ossa, ``{Comments on Conifolds},''
\href{http://dx.doi.org/10.1016/0550-3213(90)90577-Z}{{\em Nucl. Phys.} {\bf
  B342} (1990)  246--268}.

\bibitem{Ghoshal:1995wm}
D.~Ghoshal and C.~Vafa, ``{C = 1 string as the topological theory of the
  conifold},'' \href{http://dx.doi.org/10.1016/0550-3213(95)00408-K}{{\em Nucl.
  Phys.} {\bf B453} (1995)  121--128},
\href{http://arxiv.org/abs/hep-th/9506122}{{\tt arXiv:hep-th/9506122
  [hep-th]}}.

\bibitem{Giveon:1999zm}
A.~Giveon, D.~Kutasov, and O.~Pelc, ``{Holography for noncritical
  superstrings},'' \href{http://dx.doi.org/10.1088/1126-6708/1999/10/035}{{\em
  JHEP} {\bf 10} (1999)  035},
\href{http://arxiv.org/abs/hep-th/9907178}{{\tt arXiv:hep-th/9907178
  [hep-th]}}.

\bibitem{Eguchi:2004yi}
T.~Eguchi and Y.~Sugawara, ``{SL(2,R) / U(1) supercoset and elliptic genera of
  noncompact Calabi-Yau manifolds},''
  \href{http://dx.doi.org/10.1088/1126-6708/2004/05/014}{{\em JHEP} {\bf 05}
  (2004)  014},
\href{http://arxiv.org/abs/hep-th/0403193}{{\tt arXiv:hep-th/0403193
  [hep-th]}}.

\bibitem{Eguchi:2004ik}
T.~Eguchi and Y.~Sugawara, ``{Conifold type singularities, N=2 Liouville and
  SL(2:R)/U(1) theories},''
  \href{http://dx.doi.org/10.1088/1126-6708/2005/01/027}{{\em JHEP} {\bf 01}
  (2005)  027},
\href{http://arxiv.org/abs/hep-th/0411041}{{\tt arXiv:hep-th/0411041
  [hep-th]}}.

\bibitem{Ashok:2007ui}
S.~K. Ashok, R.~Benichou, and J.~Troost, ``{Non-compact Gepner Models,
  Landau-Ginzburg Orbifolds and Mirror Symmetry},''
  \href{http://dx.doi.org/10.1088/1126-6708/2008/01/050}{{\em JHEP} {\bf 01}
  (2008)  050},
\href{http://arxiv.org/abs/0710.1990}{{\tt arXiv:0710.1990 [hep-th]}}.

\bibitem{Mizoguchi:2008mk}
S.~Mizoguchi, ``{Localized Modes in Type II and Heterotic Singular Calabi-Yau
  Conformal Field Theories},''
  \href{http://dx.doi.org/10.1088/1126-6708/2008/11/022}{{\em JHEP} {\bf 11}
  (2008)  022},
\href{http://arxiv.org/abs/0808.2857}{{\tt arXiv:0808.2857 [hep-th]}}.

\bibitem{Babalic:2015fia}
E.~M. Babalic and M.~Visinescu, ``{Complete integrability of geodesic motion in
  Sasaki-Einstein toric $Y^{p,q}$ spaces},''
  \href{http://dx.doi.org/10.1142/S0217732315501801}{{\em Mod. Phys. Lett.}
  {\bf A30} (2015) no.~33, 1550180},
\href{http://arxiv.org/abs/1505.03976}{{\tt arXiv:1505.03976 [hep-th]}}.

\bibitem{Visinescu:2016xna}
M.~Visinescu, ``{Action-angle variables for geodesic motions in Sasaki-Einstein
  spaces $Y^{p,q}$},'' \href{http://dx.doi.org/10.1093/ptep/ptw172}{{\em PTEP}
  {\bf 2017} (2017) no.~1, 013A01},
\href{http://arxiv.org/abs/1611.01275}{{\tt arXiv:1611.01275 [hep-th]}}.

\bibitem{10.2307/1971373}
W.~Ballmann, M.~Brin, and P.~Eberlein, ``Structure of manifolds of nonpositive
  curvature. i,'' {\em Annals of Mathematics} {\bf 122} (1985) no.~1, 171--203.
  \url{http://www.jstor.org/stable/1971373}.

\bibitem{10.2307/1971303}
W.~Ballmann, M.~Brin, and R.~Spatzier, ``Structure of manifolds of nonpositive
  curvature. ii,'' {\em Annals of Mathematics} {\bf 122} (1985) no.~2,
  205--235. \url{http://www.jstor.org/stable/1971303}.

\bibitem{Eberlein:1996de}
P.~B. Eberlein, {\em Geometry of nonpositively curved manifolds}.
\newblock Chicago Lectures in Mathematics. University of Chicago Press,
  Chicago, IL, 1996.

\bibitem{KNIEPER2002453}
G.~Knieper, ``Chapter 6 hyperbolic dynamics and riemannian geometry,''
  \href{http://dx.doi.org/http://dx.doi.org/10.1016/S1874-575X(02)80008-X}{{\em
  Handbook of Dynamical Systems} {\bf 1} (2002)  453 -- 545}.
  \url{http://www.sciencedirect.com/science/article/pii/S1874575X0280008X}.

\bibitem{sullivan1979density}
D.~Sullivan, ``The density at infinity of a discrete group of hyperbolic
  motions,'' {\em Inst. Hautes {\'E}tudes Sci. Publ. Math} {\bf 50} (1979)
  no.~2979, 171--202.

\bibitem{patterson1976limit}
S.~J. Patterson, ``The limit set of a fuchsian group,'' {\em Acta mathematica}
  {\bf 136} (1976) no.~1, 241--273.

\bibitem{zbMATH04080312}
W.~{Ballmann}, ``{On the Dirichlet problem at infinity for manifolds of
  nonpositive curvature.},''
  \href{http://dx.doi.org/10.1515/form.1989.1.201}{{\em {Forum Math.}} {\bf 1}
  (1989) no.~2, 201--213}.

\bibitem{10.2307/120995}
G.~Knieper, ``The uniqueness of the measure of maximal entropy for geodesic
  flows on rank 1 manifolds,'' {\em Annals of Mathematics} {\bf 148} (1998)
  no.~1, 291--314. \url{http://www.jstor.org/stable/120995}.

\bibitem{sasaki1958}
S.~Sasaki, ``On the differential geometry of tangent bundles of riemannian
  manifolds,'' \href{http://dx.doi.org/10.2748/tmj/1178244668}{{\em Tohoku
  Math. J. (2)} {\bf 10} (1958) no.~3, 338--354}.
  \url{http://dx.doi.org/10.2748/tmj/1178244668}.

\bibitem{Eguchi:1978xp}
T.~Eguchi and A.~J. Hanson, ``{Asymptotically Flat Selfdual Solutions to
  Euclidean Gravity},''
\href{http://dx.doi.org/10.1016/0370-2693(78)90566-X}{{\em Phys. Lett.} {\bf
  B74} (1978)  249--251}.

\bibitem{Gibbons:1979zt}
G.~W. Gibbons and S.~W. Hawking, ``{Gravitational Multi - Instantons},''
\href{http://dx.doi.org/10.1016/0370-2693(78)90478-1}{{\em Phys. Lett.} {\bf
  B78} (1978)  430}.

\bibitem{MR992334}
P.~B. Kronheimer, ``The construction of {ALE} spaces as hyper-{K}\"ahler
  quotients,'' {\em J. Differential Geom.} {\bf 29} (1989) no.~3, 665--683.

\bibitem{Joyce:2000cm}
D.~D. Joyce, {\em Compact manifolds with special holonomy}.
\newblock Oxford Mathematical Monographs. Oxford University Press, Oxford,
  2000.

\end{thebibliography}

\providecommand{\href}[2]{#2}\begingroup\raggedright\endgroup

\end{document}